\documentclass{article}
\usepackage{color}
\usepackage{graphicx}
\usepackage{amsmath, amsthm}
\usepackage{mathtools}
\usepackage{natbib}
\usepackage{url}
\usepackage{soulutf8}
\RequirePackage[colorlinks,citecolor=blue,urlcolor=blue]{hyperref}
\usepackage{xcolor}
\usepackage{chngcntr}
\usepackage{subfig}
\usepackage{hyperref}
\usepackage{algorithm}
\usepackage[noend]{algpseudocode}
\usepackage{setspace}
\newtheorem{theorem}{Theorem}
\newtheorem{corollary}{Corollary}
\usepackage{xfrac}
\definecolor{forest}{rgb}{0.133,0.545,0.133}
\usepackage{multirow}
\usepackage{amsfonts}
\newtheorem{lemma}{Lemma}

\evensidemargin=\oddsidemargin
\usepackage{etoolbox}

\newif\ifabbreviation
\pretocmd{\thebibliography}{\abbreviationfalse}{}{}
\AtBeginDocument{\abbreviationtrue}
\DeclareRobustCommand\acroauthor[2]{%
  \ifabbreviation #2\else #1 (\mbox{#2})\fi}

\begin{document}
	\newcommand{\bb}{\boldsymbol{\beta}}

	\title{Fast Power Curve Approximation for Posterior Analyses}


	\author{Luke Hagar\footnote{Luke Hagar is the corresponding author and may be contacted at \url{lmhagar@uwaterloo.ca}.} \hspace{35pt} Nathaniel T. Stevens \bigskip \\ \textit{Department of Statistics \& Actuarial Science} \\ \textit{University of Waterloo, Waterloo, ON, Canada, N2L 3G1}}

	\date{}

	\maketitle

	\begin{abstract}

Bayesian hypothesis tests leverage posterior probabilities, Bayes factors, or credible intervals to inform data-driven decision making. We propose a framework for power curve approximation with such hypothesis tests. We present a fast approach to explore the approximate sampling distribution of posterior probabilities when the conditions for the Bernstein-von Mises theorem are satisfied.  We extend that approach to consider segments of such sampling distributions in a targeted manner for each sample size explored. These sampling distribution segments are used to construct power curves for various types of posterior analyses. Our resulting method for power curve approximation is orders of magnitude faster than conventional power curve estimation for Bayesian hypothesis tests. We also prove the consistency of the corresponding power estimates and sample size recommendations under certain conditions.
\end{abstract}

		\bigskip


	\maketitle

	\baselineskip=19.5pt


	\section{Introduction}\label{sec:intro}


 \subsection{Two-Group Hypothesis Tests}\label{sec:intro.eq}
 
	 Hypothesis tests  allow practitioners to compare scalar quantities $\theta_1$ and $\theta_2$, where the characteristic $\theta_j$ describes a comparison ($j = 1$) or reference  ($j = 2$) group. These comparisons are often facilitated using the difference between the characteristics: $\theta = \theta_1 - \theta_2$. When $\theta_1$ and $\theta_2$ are positive characteristics, these comparisons may instead be carried out using the ratio $\theta = \theta_1/\theta_2$. In Bayesian settings, hypothesis testing is facilitated via the posterior distribution of $\theta$, and point null hypotheses are rarely considered when $\theta_1$ and $\theta_2$ are continuous since there must be nonzero prior probability that such hypotheses are true \citep{gelman2013bayesian}. 
  
  This paper focuses on more plausible hypotheses of the form $H_1: \theta \in (\delta_L, \delta_U)$, where $-\infty \le \delta_L < \delta_U \le \infty$. The interval $(\delta_L, \delta_U)$  accommodates the context of comparison. We assume that larger $\theta_j$ values are preferred for illustration. For difference-based comparisons, the intervals $(\delta_L, \delta_U) = \{(0, \infty), (-\delta, \delta), (-\delta, \infty) \}$ for some $\delta > 0$ may be used to respectively assess whether $\theta_1$ is superior, practically equivalent, or noninferior to $\theta_2$ \citep{spiegelhalter1994bayesian, spiegelhalter2004bayesian}. For ratio-based comparisons, the intervals $(\delta_L, \delta_U) = \{(1, \infty), (\delta^{-1}, \delta) \}$ for some $\delta > 1$ may be used to respectively assess whether $\theta_1$ is superior or practically equivalent to $\theta_2$. 
 
 Several Bayesian interval hypothesis testing methods exist, including approaches with posterior probabilities, Bayes factors, and credible intervals. Testing methods based on posterior probabilities have been introduced in
various settings (see e.g., \citet{berry2010bayesian, brutti2014bayesian, stevens2022cpm}). Given data observed from two groups, the posterior probability $Pr(H_1 ~|~ data)$ is compared to a critical value $0.5 \le \gamma < 1$. If that probability is greater than $\gamma$, one should conclude $\theta_1 - \theta_2 \in (\delta_L, \delta_U)$. Larger values of $\gamma$ allow one to draw conclusions with more conviction. 
 
 \citet{morey2011bayes} proposed the nonoverlapping hypotheses (NOH) approach to assess the plausibility of interval hypotheses with Bayes factors \citep{jeffreys1935some, kass1995bayes}. This approach directly assigns a prior distribution to the effect size considered via the hypothesis test, which is $\theta = \theta_1 - \theta_2$ or $\theta = \theta_1/\theta_2$ in this case. The NOH Bayes factor is the ratio of the posterior odds of the complementary hypotheses $H_1:  \theta \in (\delta_L, \delta_U)$ and $H_0: \theta \notin (\delta_L, \delta_U)$ to their prior odds:
    \begin{equation}\label{eqn:bf}
	\dfrac{Pr(H_1 ~|~ data)}{1 - Pr(H_1 ~|~ data)} \div \dfrac{Pr(H_1)}{1 - Pr(H_1)}.
	\end{equation} 
The NOH Bayes factor provides support for $H_1$ over $H_0$ when its value is greater than a predetermined threshold $K \ge 1$.  

  Hypothesis testing methods with credible intervals have also been proposed \citep{gubbiotti2011bayesian, brutti2014bayesian, kruschke2018rejecting}. These methods compare the credible interval for the posterior of a univariate parameter $\theta = \theta_1 - \theta_2$ or $\theta = \theta_1/\theta_2$ to the interval $(\delta_L, \delta_U)$. A credible interval $(L_{\theta, 1 - \alpha}, U_{\theta, 1 - \alpha})$ has coverage of $1 - \alpha$ if $Pr(\theta \in (L_{\theta, 1 - \alpha}, U_{\theta, 1 - \alpha}) ~| ~ data) = 1 - \alpha$. Since this interval is not uniquely defined, the equal-tailed credible interval or highest density interval (HDI) is often considered. The equal-tailed interval is defined such that $Pr(\theta < L_{\theta, 1 - \alpha} ~| ~data) = Pr(\theta > U_{\theta, 1 - \alpha} ~| ~ data) = 1 - \alpha/2$, and the HDI  is the narrowest credible interval with coverage $1 - \alpha$. If $(L_{\theta, 1 - \alpha}, U_{\theta, 1 - \alpha})$ lies entirely within $(\delta_L, \delta_U)$, one should conclude $\theta \in (\delta_L, \delta_U)$. Posterior HDIs are not invariant to monotonic transformations, so hypothesis tests with posterior HDIs for $\theta$ and a monotonic transformation thereof may not yield the same conclusions. In contrast, hypothesis tests based on posterior probabilities, Bayes factors, and equal-tailed credible intervals are invariant to monotonic transformations that are applied to both $\theta$ and the interval endpoints.

Because posterior probabilities are used to define Bayes factors and credible intervals, this paper emphasizes hypothesis tests facilitated via posterior probabilities. However, the methods proposed in this article accommodate hypothesis tests with interval hypotheses conducted using all three overviewed methods for Bayesian inference. This broader applicability is illustrated with less emphasis throughout the paper. To ensure one has collected enough data to reliably conclude $\theta \in (\delta_L, \delta_U)$ when the corresponding hypothesis $H_1$ is true, sample size determination is an important component of the design of Bayesian hypothesis tests.

 \subsection{Bayesian Power Analysis}\label{sec:intro.bssd}

Power-based approaches to Bayesian sample size determination aim for pre-experimental probabilistic control over testing procedures for a characteristic of interest $\theta$. For two-group comparisons, $\theta = h(\theta_1, \theta_2)$ for some function $h(\cdot)$. This paper considers $h(\theta_1, \theta_2) = \theta_1 - \theta_2$ and $h(\theta_1, \theta_2) = \theta_1 /\theta_2$. In pre-experimental settings, the data have not been observed and are random variables. Data from a random sample are represented by $\boldsymbol{Y}^{_{(n)}}$, consisting of observations $\{y_{i1} \}_{i = 1}^n$ from group 1 and observations $\{y_{i2} \}_{i = 1}^{ n }$ from group 2.
 
 A \emph{design} prior $p_D(\boldsymbol{\eta})$ \citep{de2007using,berry2010bayesian,gubbiotti2011bayesian} models uncertainty regarding the model parameters $\boldsymbol{\eta} = (\boldsymbol{\eta}_1, \boldsymbol{\eta}_2)$ from each group in pre-experimental settings. The characteristic of interest $\theta_j$ for group $j$ is typically specified as a function $g(\cdot)$ of the model parameters: $\theta_j = g(\boldsymbol{\eta}_j)$ for $j = 1, 2$. Since the (informative) design prior is concentrated on parameter values that are relevant to the objective of the study, it is usually different from the \emph{analysis} prior used to analyze the observed data. The design prior gives rise to the prior predictive distribution of $\boldsymbol{Y}^{_{(n)}}$:
\begin{equation}\label{eq:prior_pred}
p(\boldsymbol{y}^{_{(n)}}) = \int \prod_{i = 1}^n f(y_{i1}; \boldsymbol{\eta}_1) \prod_{i = 1}^{ n } f(y_{i2}; \boldsymbol{\eta}_2) \hspace{1pt} p_D(\boldsymbol{\eta}) \hspace{1pt} d\boldsymbol{\eta},
\end{equation}
where $f(y; \boldsymbol{\eta}_j)$ is the model for group $j = 1, 2$. The relevant power criteria defined for the Bayesian inferential methods are considered when the data are generated from this prior predictive distribution. 

\citet{gubbiotti2011bayesian} defined two methodologies for specifying (\ref{eq:prior_pred}): the conditional and predictive approaches. The conditional approach uses a degenerate design prior 
$p_D(\boldsymbol{\eta})$ that assigns a probability of 1 to fixed \emph{design values} $\boldsymbol{\eta}_{1,0}$ and $\boldsymbol{\eta}_{2,0}$ for the model parameters. These fixed values characterize the data generation process for all hypothetical repetitions of the testing procedure as in frequentist sample size calculations. The predictive approach uses a nondegenerate $p_D(\boldsymbol{\eta})$. For each hypothetical repetition of the testing procedure, different design values $\boldsymbol{\eta}_{0} = (\boldsymbol{\eta}_{1,0}, \boldsymbol{\eta}_{2,0}) \sim p_D(\boldsymbol{\eta})$ characterize the data generation process, which is arguably more consistent with the Bayesian framework. However, the cognitive complexity associated with specifying nondegenerate design priors may be prohibitive for certain practitioners. This paper therefore considers both the conditional and predictive approaches. 





Bayesian methods for power analysis have been proposed in a variety of contexts \citep{berry2010bayesian, gubbiotti2011bayesian, brutti2014bayesian}. In these contexts, one aims to select a sample size $n$ to ensure the probability of correctly concluding that $H_1 : \theta \in (\delta_L, \delta_U)$ is true is at least $\Gamma$ for some target power $\Gamma \in (0,1)$. For hypothesis tests with posterior probabilities, the selected sample size ensures that
\begin{equation}\label{eq:power_prob}
\mathbb{E}\left[\mathbb{I}\{Pr(H_1 \hspace{1pt}| \hspace{1pt} \boldsymbol{Y}^{_{(n)}}) \ge \gamma\}\right] \ge \Gamma,
\end{equation}
for some critical value $\gamma \in [0.5,1)$. To \emph{correctly} conclude that $H_1$ is true in (\ref{eq:power_prob}), we must choose a design prior $p_{D}(\boldsymbol{\eta})$ such that $p_{D}(H_1) = 1$. Power analyses for hypothesis tests with Bayes factors are related. It follows from (\ref{eqn:bf}) that the NOH Bayes factor exceeds $K$ if and only if
  \begin{equation}\label{eqn:bf_pp}
	Pr(H_1~| ~ data) > \dfrac{K \times Pr(H_1)}{ 1 - (K - 1) \times Pr(H_1)} .
	\end{equation} 
The power criterion for NOH Bayes factors with threshold $K \ge 1$ is therefore a special case of (\ref{eq:power_prob}) when the critical value $\gamma$ equals the right side of (\ref{eqn:bf_pp}). 

For hypothesis tests with credible intervals, the quantity in (\ref{eq:power_prob}) is replaced with 
\begin{equation}\label{eq:power_hdi}
\mathbb{E}\left[\mathbb{I}\{L_{\theta, 1 - \alpha}(\boldsymbol{Y}^{_{(n)}}) > \delta_L  ~~\cap~~ U_{\theta, 1 - \alpha}(\boldsymbol{Y}^{_{(n)}}) < \delta_U\}\right] \ge \Gamma,
\end{equation}
where the endpoints of the credible interval are henceforth denoted $L_{\theta, 1 - \alpha}(\boldsymbol{Y}^{_{(n)}})$ and $U_{\theta, 1 - \alpha}(\boldsymbol{Y}^{_{(n)}})$ to emphasize their dependence on the data. If the posterior credible interval is equal tailed, the power criterion in (\ref{eq:power_hdi}) simplifies to
\begin{equation}\label{eq:power_ci}
\mathbb{E}\left[\mathbb{I}\{Pr(\theta < \delta_L | \hspace{1pt} \boldsymbol{Y}^{_{(n)}}) < \alpha/2  ~~\cap~~ Pr(\theta > \delta_U | \hspace{1pt} \boldsymbol{Y}^{_{(n)}}) < \alpha/2\}\right] \ge \Gamma.
\end{equation}
When $1-\alpha = \gamma$, (\ref{eq:power_hdi}) and (\ref{eq:power_ci}) impose stricter criteria than (\ref{eq:power_prob}). At least $100\times\gamma\%$ of the posterior for $\theta$ must lie within the interval $(\delta_L, \delta_U)$ for $(L_{\theta, 1 - \alpha}(\boldsymbol{Y}^{_{(n)}}), U_{\theta, 1 - \alpha}(\boldsymbol{Y}^{_{(n)}}))$ to also be contained in this interval. The plot of the quantity in (\ref{eq:power_prob}), (\ref{eq:power_hdi}), or (\ref{eq:power_ci}) as a function of the sample size $n$ is called the power curve.

To support flexible study design, sample sizes that satisfy power criteria can be found using simulation. Most simulation-based procedures for power analysis with design priors follow a similar process \citep{wang2002simulation}. First, a sample size $n$ is selected. Second, a value $\boldsymbol{\eta}_0 = (\boldsymbol{\eta}_{1,0}, \boldsymbol{\eta}_{2,0})$ is drawn from the (perhaps degenerate) design prior $p_D(\boldsymbol{\eta})$. Third, data $\boldsymbol{y}^{_{(n)}}$ are generated according to the models $f(y; \boldsymbol{\eta}_{1,0})$ and $f(y; \boldsymbol{\eta}_{2,0})$. Fourth, the posterior of $\theta$ given $\boldsymbol{y}^{_{(n)}}$ is approximated to compute $Pr(H_1  \hspace{1pt}| \hspace{1pt}\boldsymbol{y}^{_{(n)}})$. This process is repeated many times to estimate a sampling distribution of posterior probabilities, which is used to determine whether the power criterion is satisfied with probability $\Gamma$ for the selected sample size $n$. 

These simulation-based approaches can be very computationally intensive as many posteriors must be approximated to estimate the sampling distribution for each sample size $n$ considered. \citet{wang2002simulation} recommended using bisection methods or grid searches to streamline the exploration of sample sizes. Yet even when such methods circumvent the need for practitioners to choose which sample sizes $n$ to explore, time is still wasted considering sample sizes that are excessively large or much too small to satisfy the power criterion. This computational inefficiency is compounded over all combinations of the design inputs that practitioners wish to consider when designing Bayesian hypothesis tests -- including the critical value $\gamma$, interval $(\delta_L, \delta_U)$, target power $\Gamma$, and design and analysis priors. A fast framework for power curve approximation with posterior analyses based on sampling distribution segments would mitigate this issue and expedite study design. 

 \subsection{Contributions}\label{sec:intro.cont}

The remainder of this article is structured as follows. We describe a food expenditure example involving the comparison of gamma tail probabilities in Section \ref{sec:ex.1}. This example is referenced throughout the paper to motivate the proposed methods. In Section \ref{sec:sampling}, we propose a method to map the sampling distribution of posterior probabilities to low-dimensional hypercubes. We also prove that the resulting approximation to the sampling distribution gives rise to consistent power estimates under certain conditions. In Section \ref{sec:curve}, we exploit this mapping to quickly approximate power curves with low-discrepancy sequences. This approach is fast because for a given sample size, we consider only a segment of the approximate sampling distribution of posterior probabilities. Even without estimating entire sampling distributions, this method prompts consistent sample size recommendations. In Section \ref{sec:study}, we conduct numerical studies to explore the performance of our power curve approximation method in several settings. We conclude with a discussion of extensions to this work in Section \ref{sec:conclusion}.

	\section{Motivating Example with Gamma Tail Probabilities}\label{sec:ex.1}

\begin{figure}[!t] \centering 
		\includegraphics[width = 0.975\textwidth]{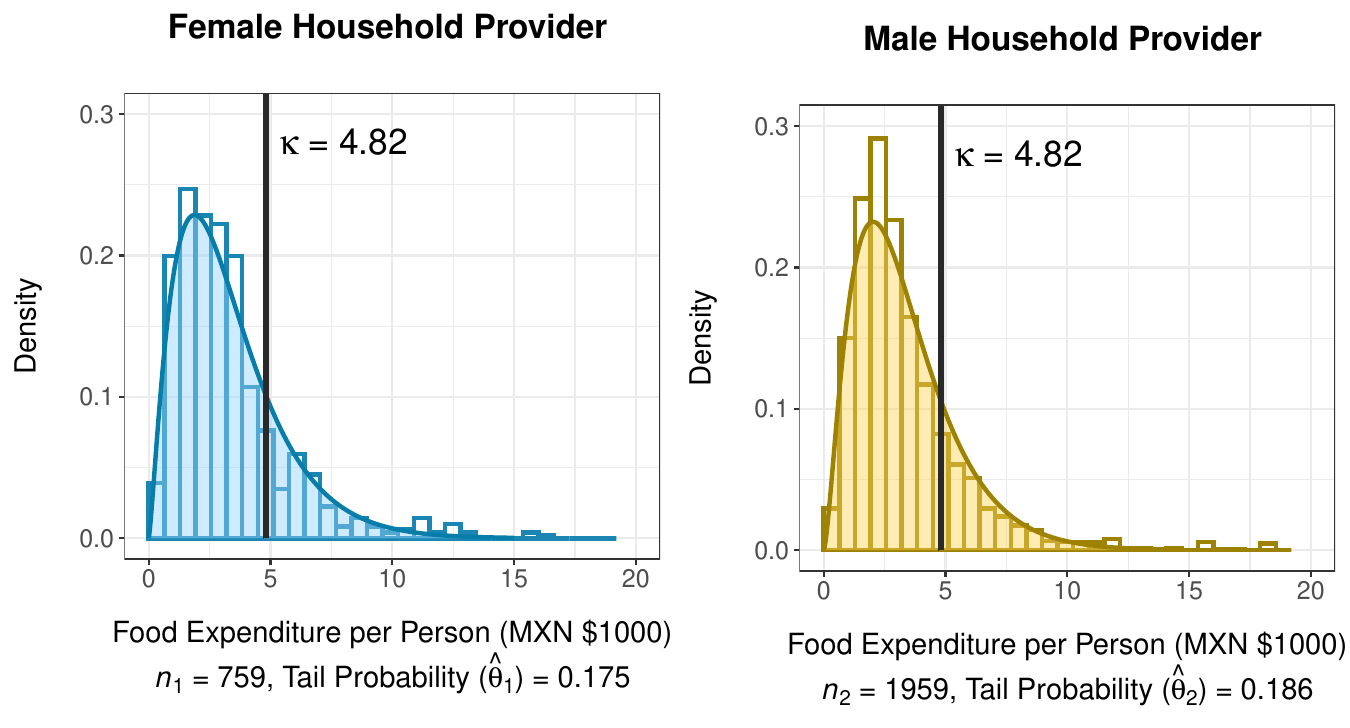} 
		\caption{\label{fig:cpm_gamma} Distribution of quarterly food expenditure per person in each group.} 
	\end{figure}

	Mexico's National Institute of Statistics and Geography conducts a biennial survey to monitor household income and expenses along with sociodemographic characteristics. We refer to this survey by its Spanish acronym ENIGH. In the ENIGH 2020 survey \citep{enigh2020}, each surveyed household was assigned a socioeconomic class: lower, lower-middle, upper-middle, and upper. We use data from the lower-middle income households (the most populous class) in the Mexican state of Aguascalientes. We split the households into two groups based on the sex of the household's main provider. Each household has a weighting factor used to include its observation between one and four times in our data set. The datum $y_{ij}$ collected for each household $i = 1,...,n_j, ~j = 1,2$ is its quarterly expenditure on food per person measured in thousands of Mexican pesos (MXN \$1000). We exclude the 0.41\% of households that report zero quarterly expenditure on food to accommodate the gamma model's positive support. This respectively yields $n_1 = 759$ and $n_2 = 1959$ observations in the female ($j = 1$) and male ($j = 2$) provider groups that are visualized in Figure \ref{fig:cpm_gamma}.

 Here, we compare tail probabilities for each distribution such that $\theta_j = Pr(y_{ij} > \kappa)$, where $\kappa$ is a scalar value from the support of distribution $j = 1,2$. The threshold of $\kappa = 4.82$ for this example is the median quarterly food expenditure per person (in MXN \$1000) for \textit{upper} income households in Aguascalientes after accounting for weighting factors. We will later use the ratio $\theta_1 /\theta_2$ to compare the probabilities that lower-middle income households with female and male providers spend at least as much on food per person as the typical upper income household.  The observed proportions of households that spend at least \$4820 MXN on food per person are $\hat\theta_1 = 0.175$ and $\hat\theta_2 = 0.186$. We assign uninformative $\text{GAMMA}(2, 0.25)$ priors to both the shape $\alpha_j$ and rate $\beta_j$ parameters of the gamma model for group $j = 1, 2$. We let $\boldsymbol{\eta}_j = (\alpha_j, \beta_j)$ for $j = 1, 2$. We obtain $10^5$ posterior draws for $\boldsymbol{\eta}_1$ and $\boldsymbol{\eta}_2$ using Markov chain Monte Carlo (MCMC) methods. The gamma distributions characterized by the posterior means for $\boldsymbol{\eta}_1$ and $\boldsymbol{\eta}_2$ are superimposed on the histograms in Figure \ref{fig:cpm_gamma}. We use this example in subsequent sections of this paper to provide a context to motivate Bayesian hypothesis testing.
		


 \section{The Sampling Distribution of Posterior Probabilities}\label{sec:sampling}

 \subsection{Analytical Approximations to the Posterior}\label{sec:samp.norm}

  Traditional approaches to power curve approximation for posterior analyses require that we estimate the sampling distribution of posterior probabilities for various sample sizes $n$. These approaches are slow because we wastefully estimate \emph{entire} sampling distributions for sample sizes that are much too large or small when searching for a suitable sample size $n$. To map posterior probabilities to low-dimensional hypercubes and explore segments of sampling distributions, we leverage normal approximations to the posterior of $\theta$ based on limiting results. In this subsection, we overview several analytical posterior approximation methods and conditions that must hold for these approximations to be suitable. 
  
 The first normal approximation to the posterior that we consider follows from the Bernstein-von Mises (BvM) theorem \citep{vaart1998bvm}. We now describe how our framework for power curve approximation satisfies the conditions for the BvM theorem. Our framework assumes that data $\{y_{i1}\}_{i = 1}^{n}$ and $\{y_{i2}\}_{i = 1}^{n}$ are to be collected independently, where the data generation process for group $j$ is characterized by the model $f(y; \boldsymbol{\eta}_{j,0})$ parameterized by $\boldsymbol{\eta}_{j} \in \mathbb{R}^d$. Here, $\boldsymbol{\eta}_{1,0}$ and $\boldsymbol{\eta}_{2,0}$ are design values for the distributional parameter(s) that are drawn according to the design prior $p_D(\boldsymbol{\eta})$. The models $f(y; \boldsymbol{\eta}_{1,0})$ and $f(y; \boldsymbol{\eta}_{2,0})$ are herein referred to as \textit{design} distributions. If the predictive approach is used to specify (\ref{eq:prior_pred}), then the design distributions differ for each hypothetical repetition of the hypothesis test.

 The conditions in the following two paragraphs must be satisfied for all $\boldsymbol{\eta}_{0} = (\boldsymbol{\eta}_{1,0}, \boldsymbol{\eta}_{2,0}) \sim p_D(\boldsymbol{\eta})$. This requirement should be taken into account when specifying $p_D(\boldsymbol{\eta})$. We emphasize that the design values $\boldsymbol{\eta}_{j,0}$ are different from the random variables $\boldsymbol{\eta}_{j}$ that parameterize the model for groups $j = 1,2$ in Bayesian settings. When specifying the models $f(y; \boldsymbol{\eta}_{j,0})$, we also specify fixed values $\theta_{j,0}$ for the random variables $\theta_j$: $\theta_{j,0} = g(\boldsymbol{\eta}_{j,0})$. We require that $g(\boldsymbol{\eta}_{j})$ is differentiable at $\boldsymbol{\eta}_{j} = \boldsymbol{\eta}_{j,0}$ for $j = 1, 2$. A fixed value for the univariate characteristic $\theta_0 = h(\theta_{1,0}, \theta_{2,0})$ is also specified, where $h(\theta_1, \theta_2)$ is a differentiable function at $\theta_1 = \theta_{1,0}$ and $\theta_2 = \theta_{2,0}$. These derivatives of $g(\cdot)$ and $h(\cdot)$ must be nonzero.  
         
         The four assumptions that must be satisfied to invoke the BvM theorem \citep{vaart1998bvm} are detailed in Appendix A.1 of the supplement \citep{hagar2023supp}. The first three assumptions involve the models $f(y; \boldsymbol{\eta}_{1,0})$ and $f(y; \boldsymbol{\eta}_{2,0})$; they are weaker than the regularity conditions for the asymptotic normality of the maximum likelihood estimator (MLE) \citep{lehmann1998theory}, which are listed in Appendix A.2 of the supplement. The final assumption for the BvM theorem regards prior specification for the random variables. For our purposes, $\boldsymbol{\eta}_1$ and $\boldsymbol{\eta}_2$ are the random variables for which we explicitly or implicitly assign prior distributions. We require that the (analysis) prior distribution of $\boldsymbol{\eta}_j$ be continuous in a neighbourhood of $\boldsymbol{\eta}_{j,0}$ with positive density at $\boldsymbol{\eta}_{j,0}$ for $j = 1,2$. This condition ensures that the posterior of $\theta$ converges to a neighbourhood of $\theta_0 = h(g(\boldsymbol{\eta}_{1,0}), g(\boldsymbol{\eta}_{2,0}))$. This convergence is required for our method for power curve approximation introduced in Section \ref{sec:curve}. 

         Under the conditions for the BvM theorem, the posterior of $\theta$ converges to the $\mathcal{N}(\theta_0,\mathcal{I}(\theta_0)^{-1}/n)$ distribution in the limit of infinite data \citep{vaart1998bvm}, where $\mathcal{I}(\theta_0)$ is the Fisher information for $\theta$ evaluated at $\theta_0$. In practice, $\theta_0$ is estimated from the to-be-observed data $\boldsymbol{Y}^{_{(n)}}$ using the MLE $\hat\theta_n$ or the posterior mode $\tilde\theta_n$. In the limiting case, it does not matter which estimator for $\theta$ is used because both $\hat\theta_n$ and $\tilde\theta_n$ converge in probability to $\theta_0$ when the conditions for the BvM theorem are satisfied. We use the MLE instead of the posterior mode for reasons discussed in Section \ref{samp:dim}. We therefore consider the following normal distribution based on the BvM theorem as one option to approximate the posterior of $\theta$: 
         \begin{equation}\label{eqn:bvm_mle}
\mathcal{N}\left( h(g(\hat{\boldsymbol{\eta}}_{1,n}),g(\hat{\boldsymbol{\eta}}_{2,n})), \dfrac{1}{n}\sum_{j = 1}^2\left[\dfrac{\partial h}{\partial\theta_j}\right]^2_{\theta_j = g(\hat{ \boldsymbol{\eta}}_{j,n})} \left[\dfrac{\partial g}{\partial\boldsymbol{\eta}}^T \mathcal{I}(\boldsymbol{\eta})^{-1} \dfrac{\partial g}{\partial\boldsymbol{\eta}}\right]_{\boldsymbol{\eta} = \hat{ \boldsymbol{\eta}}_{j,n}}\right).
	\end{equation} 
Given $\mathcal{I}(\boldsymbol{\eta})^{-1}$, the multivariate delta method prompts the result in (\ref{eqn:bvm_mle}) since $\hat\theta_n = h(g(\hat{\boldsymbol{\eta}}_{1,n}), g(\hat{\boldsymbol{\eta}}_{2,n}))$ is a function of the MLEs $\hat{\boldsymbol{\eta}}_{1,n}$ and $\hat{\boldsymbol{\eta}}_{2,n}$. We note that the variance in (\ref{eqn:bvm_mle}) is $\mathcal{I}(\hat\theta_n)^{-1}/n$.  While it does not account for the priors, the approximation in (\ref{eqn:bvm_mle}) is useful because it prompts theoretical results about the limiting behaviour of the sampling distribution of posterior probabilities in Section \ref{samp:theo}.

          We also consider the Laplace approximation to the posterior of $\theta$ that \emph{does} account for the priors. This is useful when the sample size $n$ is large enough to ensure the posterior is approximately normal but not large enough to guarantee the relevant priors have no substantial impact on the posterior mean and variance. For groups $j = 1$ and 2, the Laplace approximation is based on the Taylor series expansion of $\log(p_j(\boldsymbol{\eta}_j\hspace{1pt}|\hspace{1pt} data))$ centered at the posterior mode $\tilde{ \boldsymbol{\eta}}_{j,n} = \text{arg\,max}_{{\boldsymbol{\eta}_j}} ~p_j(\boldsymbol{\eta}_j\hspace{1pt}|\hspace{1pt} data)$ \citep{gelman2013bayesian}. The multivariate delta method prompts the following normal approximation to the posterior of $\theta$ that accounts for the analysis priors $p_1(\boldsymbol{\eta}_1)$ and $p_2(\boldsymbol{\eta}_2)$:
                   \begin{equation}\label{eqn:norm}
\mathcal{N}\left( h(g(\tilde{\boldsymbol{\eta}}_{1,n}),g(\tilde{\boldsymbol{\eta}}_{2,n})), \sum_{j = 1}^2\left[\dfrac{\partial h}{\partial\theta_j}\right]^2_{\theta_j = g(\tilde{ \boldsymbol{\eta}}_{j,n})} \left[\dfrac{\partial g}{\partial\boldsymbol{\eta}}^T \mathcal{J}_j(\boldsymbol{\eta})^{-1} \dfrac{\partial g}{\partial\boldsymbol{\eta}}\right]_{\boldsymbol{\eta} = \tilde{ \boldsymbol{\eta}}_{j,n}}\right),
	\end{equation} 
\begin{equation*}\label{eqn:norm2}
\text{where}~~~~\mathcal{J}_j(\boldsymbol{\eta}) = -\dfrac{\partial^2}{\partial\boldsymbol{\eta}^2} \log(p_j(\boldsymbol{\eta}|\hspace{1pt} data)).
	\end{equation*} 
 We generally recommend using Laplace approximations with our framework for power curve approximation, but the approximation in (\ref{eqn:norm}) is computationally burdensome and suboptimal in certain situations as explained in Section \ref{samp:data}. 

 \subsection{Mapping Posteriors to Low-Dimensional Hypercubes}\label{samp:dim}

 We next propose methods to map posteriors to low-dimensional hypercubes when using normal approximations to the posterior that do not and do account for the priors in this subsection and Section \ref{samp:data}, respectively. The methods presented in the remainder of this section allow us to consider sampling distribution segments in Section \ref{sec:curve}. To generate samples of size $n$ from each design distribution $f(y; \boldsymbol{\eta}_{1,0})$ and $f(y; \boldsymbol{\eta}_{2,0})$, one typically uses a pseudorandom sequence $\boldsymbol{u}_1,\boldsymbol{u}_2, ..., \boldsymbol{u}_m \in [0,1]^{2 n}$ with length $m$. However, the approximation in (\ref{eqn:bvm_mle}) does not directly use the observations from the generated sample $\boldsymbol{y}^{_{(n)}}$. Instead, $\boldsymbol{y}^{_{(n)}}$ is used to compute maximum likelihood estimates $\hat{\boldsymbol{\eta}}_{1,n} \in \mathbb{R}^d$ and $\hat{\boldsymbol{\eta}}_{2,n} \in \mathbb{R}^d$, which yield $\hat{\theta}_n = h(g(\hat{\boldsymbol{\eta}}_{1,n}),g(\hat{\boldsymbol{\eta}}_{2,n}))$. As such, we do not need to simulate data $\boldsymbol{Y}^{_{(n)}}$ from the prior predictive distribution for a given sample size $n$. We instead recommend simulating from the approximate distributions for the MLEs of ${\boldsymbol{\eta}}_{1}$ and ${\boldsymbol{\eta}}_{2}$. This reduces the dimension of the simulation from $2n$ to $2d$ (since $d << n$).

 For sufficiently large $n$, the MLEs $\hat{\boldsymbol{\eta}}_{j,n}$ for groups $j = 1,2$ approximately and independently follow $\mathcal{N}(\boldsymbol{\eta}_{j,0}, \mathcal{I}(\boldsymbol{\eta}_{j,0})^{-1}/n)$ distributions. The MLE -- and not the posterior mode -- is used with the approximation in (\ref{eqn:bvm_mle}) because we can easily simulate from its limiting distribution. Both $\hat{\boldsymbol{\eta}}_{1, n}$ and $\hat{\boldsymbol{\eta}}_{2, n}$ have dimension $d$, so their joint limiting distribution has dimension $2 d$. When using pseudorandom number generation, we now require a sequence $\boldsymbol{u}_1,\boldsymbol{u}_2, ..., \boldsymbol{u}_{m} \in [0,1]^{2 d}$. Algorithm \ref{alg1} details how we map a single point $\boldsymbol{u} = (u_1, u_2, ..., u_{2d}) \in [0,1]^{2 d}$ to the posterior approximation based on the BvM theorem in (\ref{eqn:bvm_mle}). The square brackets in the subscripts denote the indices of the relevant subvectors and submatrices.  

\begin{algorithm}
\caption{Mapping Posteriors to $[0,1]^{2d}$ with the BvM Theorem}
\label{alg1}

\begin{algorithmic}[1]
\setstretch{1}
\Procedure{MapBvM}{$f(y; \boldsymbol{\eta})$, $\boldsymbol{\eta}_{1,0}$, $\boldsymbol{\eta}_{2,0}$, $g(\cdot)$, $h(\cdot)$, $n$, $\boldsymbol{u}$}
\For{$j$ in 1:2}
\State $\boldsymbol{\mu} \leftarrow \boldsymbol{\eta}_{j,0}; \boldsymbol{\Sigma} \leftarrow \mathcal{I}(\boldsymbol{\eta}_{j,0})^{-1}/n$.
\State Generate $\hat{\boldsymbol{\eta}}_{j,n}(\boldsymbol{u})_{[1]}$ as the $u_{(j-1)d + 1}$-quantile of the $\mathcal{N}(\boldsymbol{\mu}_{[1]}, \boldsymbol{\Sigma}_{[1,1]})$ CDF.
\For{$k$ in 1:$(d-1)$}
    \State Generate $\hat{\boldsymbol{\eta}}_{j,n}(\boldsymbol{u})_{[k+1]}$ as the $u_{(j-1)d + (k+1)}$-quantile of the normal CDF  with mean $\boldsymbol{\mu}_{[k+1]} +$ \linebreak \hspace*{42pt} $\boldsymbol{\Sigma}_{[1,1:k]}\boldsymbol{\Sigma}^{-1}_{[1:k,1:k]}(\hat{\boldsymbol{\eta}}_{j,n}(\boldsymbol{u})_{[1:k]} -\boldsymbol{\mu}_{[1:k]})$ and  variance $\boldsymbol{\Sigma}_{[k+1,k+1]} - \boldsymbol{\Sigma}_{[1,1:k]}\boldsymbol{\Sigma}^{-1}_{[1:k,1:k]}\boldsymbol{\Sigma}_{[1:k,1]}$.
    \EndFor
    \EndFor
    \State Use $\hat{\boldsymbol{\eta}}_{1,n}(\boldsymbol{u})$, $ \hat{\boldsymbol{\eta}}_{2,n}(\boldsymbol{u})$, and the partial derivatives of $g(\cdot)$ and $h(\cdot)$ to obtain (\ref{eqn:bvm_mle}).
\EndProcedure

\end{algorithmic}
\end{algorithm}

The posterior approximation in (\ref{eqn:bvm_mle}) depends on the model $f(y; \boldsymbol{\eta})$, the design values $\boldsymbol{\eta}_{1,0}$ and $\boldsymbol{\eta}_{2,0}$ drawn from $p_D(\boldsymbol{\eta})$, the functions $g(\cdot)$ and $h(\cdot)$, and the maximum likelihood estimates $\hat{\boldsymbol{\eta}}_{1, n}$ and $\hat{\boldsymbol{\eta}}_{2, n}$. The model $f(y; \boldsymbol{\eta})$ determines the Fisher information matrix $\mathcal{I}(\boldsymbol{\eta})$. Given the design values, Lines 2 to 6 of Algorithm \ref{alg1} generate $\hat{\boldsymbol{\eta}}_{j, n} \sim \mathcal{N}(\boldsymbol{\eta}_{j,0}, \mathcal{I}(\boldsymbol{\eta}_{j,0})^{-1}/n)$ for $j = 1, 2$ using cumulative distribution function (CDF) inversion with conditional univariate normal distributions. The first $d$ coordinates of $\boldsymbol{u} \in [0,1]^{2 d}$ prompt $\hat{\boldsymbol{\eta}}_{1, n}$, and the final $d$ coordinates give rise to $\hat{\boldsymbol{\eta}}_{2, n}$. These maximum likelihood estimates can be substituted into (\ref{eqn:bvm_mle}) along with the partial derivatives of $g(\cdot)$ and $h(\cdot)$ to approximate the posterior of $\theta$. 

In practice, we may require fewer observations for the sampling distributions of the MLEs to be approximately normal if we consider some transformation of ${\boldsymbol{\eta}}_{j}$. For  the gamma model, both parameters in ${\boldsymbol{\eta}}_{j} = (\alpha_j, \beta_j)$ must be positive, but the $\mathcal{N}(\boldsymbol{\eta}_{j,0}, \mathcal{I}(\boldsymbol{\eta}_{j,0})^{-1}/n)$ distribution could admit nonpositive values for small $n$. To obtain a sample of positive $\hat{\boldsymbol{\eta}}_{1, n}$ and $\hat{\boldsymbol{\eta}}_{2, n}$ values for any sample size $n$ with the gamma model, we exponentiate a sample of approximately normal MLEs of $\log({\boldsymbol{\eta}}_{1})$ and $\log({\boldsymbol{\eta}}_{2})$. For an arbitrary model, appropriate transformations could similarly be applied to any parameters in ${\boldsymbol{\eta}}_{j}$ that do not have support on $\mathbb{R}$. Similarly, the posterior of a monotonic transformation of $\theta$ may need to be considered for the normal approximation in (\ref{eqn:bvm_mle}) to be suitable for moderate $n$. For instance, the posterior of $\log(\theta_1) - \log(\theta_2)$ may be better approximated by a normal distribution than that of $\theta_1 /\theta_2$. Rather than introduce new notation for these untransformed and transformed variables, we assume that ${\boldsymbol{\eta}}_{1}$, ${\boldsymbol{\eta}}_{2}$, and $\theta$ are specified to improve the quality of the relevant normal approximations in (\ref{eqn:bvm_mle}) and (\ref{eqn:norm}). Because priors are typically specified for ${\boldsymbol{\eta}}_{1}$ and  ${\boldsymbol{\eta}}_{2}$ before making such transformations, relevant Jacobians must be considered when using normal approximations to the posterior that account for the prior distributions.

 \subsection{Mapping Posteriors with Prior Information}\label{samp:data}

The method for mapping posteriors to $[0,1]^{2d}$ proposed in Section \ref{samp:dim} does not account for the prior distributions. The Laplace approximation to the posterior of $\theta$ in (\ref{eqn:norm}) that accounts for the priors requires an observed sample $\boldsymbol{y}^{_{(n)}}$ -- not just the maximum likelihood estimates  $\hat{\boldsymbol{\eta}}_{1,n}$ and $ \hat{\boldsymbol{\eta}}_{2,n}$. In this subsection, we present two methods for posterior mapping that account for the priors. The first method is ideal when the design distributions belong to the exponential family \citep{lehmann1998theory}, whereas the second method allows for more flexibility when specifying the models $f(y; \boldsymbol{\eta}_j)$.

 When the design distributions belong to the exponential family, the relevant probability mass or density function takes the form
                   \begin{equation*}\label{eqn:exp_fam}
f(y; \boldsymbol{\eta}_j) = \text{exp}\left[ \sum_{s = 1}^d C_s(\boldsymbol{\eta}_j)T_s(y) - A(\boldsymbol{\eta}_j) + B(y) \right],
	\end{equation*} 
 where $A(\boldsymbol{\eta}_j)$, $B(y)$, $C_s(\boldsymbol{\eta}_j)$, and $T_s(y)$ are known functions for $s = 1, ..., d$. For group $j$, $T_{j^{\dagger}}(\boldsymbol{y}^{_{(n)}}) = (\sum_{i = 1}^nT_1(y_{ij}), ..., \sum_{i = 1}^nT_d(y_{ij}))$ are called sufficient statistics that provide as much information about the parameter $\boldsymbol{\eta}_j$ as the entire sample. The first derivative of the log-likelihood with respect to the $k^{\text{th}}$ component of $\boldsymbol{\eta}_j$ is then
   \begin{equation}\label{eqn:exp_deriv}
\dfrac{\partial}{\partial \boldsymbol{\eta}_{j [k]}} l(\boldsymbol{\eta}_j; \boldsymbol{y}^{_{(n)}}) = -n\dfrac{\partial }{\partial \boldsymbol{\eta}_{j [k]}}A(\boldsymbol{\eta}_j) + \sum_{s = 1}^d \dfrac{\partial }{\partial \boldsymbol{\eta}_{j [k]}}C_s(\boldsymbol{\eta}_j) \sum_{i = 1}^n T_s(y_{ij}).
	\end{equation} 
At the maximum likelihood estimate $\hat{\boldsymbol{\eta}}_{j,n}$ for an observed sample $\boldsymbol{y}^{_{(n)}}$, all $d$ partial derivatives in (\ref{eqn:exp_deriv}) equal 0. A $d$-parameter model in the exponential family has $d$ nonredundant sufficient statistics, so all components of $T_{j^{\dagger}}(\boldsymbol{y}^{_{(n)}})$ can be recovered by substituting the maximum likelihood estimate $\hat{\boldsymbol{\eta}}_{j,n}$ into the system of linear equations in (\ref{eqn:exp_deriv}).  Algorithm \ref{alg2} details how we map a single point $\boldsymbol{u} \in [0,1]^{2d}$ to the posterior approximation in (\ref{eqn:norm}) based on Laplace's method. For models $f(y; \boldsymbol{\eta}_{1})$ and $f(y; \boldsymbol{\eta}_{2})$ in the exponential family, we emphasize that $p_j(\boldsymbol{\eta}_j|\hspace{1pt} data) = p_j(\boldsymbol{\eta}_j|\hspace{1pt} T_{j^{\dagger}}(\boldsymbol{y}^{_{(n)}}))$ for $j = 1, 2$. The sufficient statistics $T_{j^{\dagger}}(\boldsymbol{y}^{_{(n)}})$ returned in Line 6 therefore prompt $\mathcal{J}_j(\boldsymbol{\eta})$ in (\ref{eqn:norm}). 

\begin{algorithm}
\caption{Mapping Posteriors to $[0,1]^{2d}$ with Laplace's Method}
\label{alg2}

\begin{algorithmic}[1]
\setstretch{1}
\Procedure{MapLaplace}{$f(y; \boldsymbol{\eta})$, $\boldsymbol{\eta}_{1,0}$, $\boldsymbol{\eta}_{2,0}$, $g(\cdot)$, $h(\cdot)$, $n$, $\boldsymbol{u}$, $p_1(\boldsymbol{\eta}_1)$, $p_2(\boldsymbol{\eta}_2)$}
\State Generate $\hat{\boldsymbol{\eta}}_{1,n}(\boldsymbol{u})$ and $ \hat{\boldsymbol{\eta}}_{2,n}(\boldsymbol{u})$ using Lines 2 to 6 of Algorithm \ref{alg1}.
\For{$j$ in 1:2}
    \State Equate the system of equations in (\ref{eqn:exp_deriv}) to 0 with $\boldsymbol{\eta}_j = \hat{\boldsymbol{\eta}}_{j,n}(\boldsymbol{u})$ to solve for $T_{j^{\dagger}}(\boldsymbol{y}^{_{(n)}})$.
    \State Use $T_{j^{\dagger}}(\boldsymbol{y}^{_{(n)}})$ to obtain the posterior mode $\tilde{\boldsymbol{\eta}}_{j,n}$ via optimization.
    \EndFor
    \State Use $\tilde{\boldsymbol{\eta}}_{1,n}(\boldsymbol{u})$, $ \tilde{\boldsymbol{\eta}}_{2,n}(\boldsymbol{u})$, $T_{1^{\dagger}}(\boldsymbol{y}^{_{(n)}})$, and $T_{2^{\dagger}}(\boldsymbol{y}^{_{(n)}})$ along with the partial  derivatives  of $g(\cdot)$ and $h(\cdot)$ to \linebreak \hspace*{12pt} obtain (\ref{eqn:norm}).
\EndProcedure

\end{algorithmic}
\end{algorithm}

We note that Algorithm \ref{alg2} may not provide a serviceable approach when the design distributions are not members of the exponential family. For instance, this method could not be applied if a Weibull model were chosen for the motivating example in Section \ref{sec:ex.1} in lieu of the gamma distribution. When the shape parameter of the Weibull distribution is unknown, its minimal sufficient statistic consists of the entire sample: $\{y_{ij} \}_{i = 1}^n$ for $j = 1,2$. We therefore develop a hybrid approach to posterior mapping that accounts for the priors when low-dimensional sufficient statistics cannot be recovered from the maximum likelihood estimates $\hat{\boldsymbol{\eta}}_{1,n}$ and $\hat{\boldsymbol{\eta}}_{2,n}$. 

This hybrid approach leverages the following result, which holds true when $\boldsymbol{\eta}_j \approx \hat{\boldsymbol{\eta}}_{j,n}$ for sufficiently large $n$: 
 \begin{equation}\label{eqn:opt_exp}
\begin{split}
\log(p_j(\boldsymbol{\eta}_j|\hspace{1pt} \boldsymbol{y}^{_{(n)}})) \approx l(\hat{\boldsymbol{\eta}}_{j,n};\hspace{1pt} \boldsymbol{y}^{_{(n)}}) - \frac{n}{2}(\boldsymbol{\eta}_j - \hat{\boldsymbol{\eta}}_{j,n})^T\mathcal{I}(\hat{\boldsymbol{\eta}}_{j,n})(\boldsymbol{\eta}_j - \hat{\boldsymbol{\eta}}_{j,n}) + \log(p_j(\boldsymbol{\eta}_j)).
\end{split}
	\end{equation} 
This result follows from the second-order Taylor approximation to the log-posterior of $\boldsymbol{\eta}_j$ around $\hat{\boldsymbol{\eta}}_{j,n}$, where the observed information is replaced with the (expected) Fisher information. The approximation to the log-likelihood function does not have a first-order term because the score function is 0 at $\hat{\boldsymbol{\eta}}_{j,n}$. We note that although the first term on the right side of (\ref{eqn:opt_exp}) depends on the data $\boldsymbol{y}^{_{(n)}}$, it is a constant. An approximation to the posterior mode is the value that maximizes the right side of (\ref{eqn:opt_exp}): $\boldsymbol{\eta}_{j,n}^{*}$. We consider the following normal approximation to the posterior of $\theta$:
                    \begin{equation}\label{eqn:norm3}
\mathcal{N}\left( h(g(\boldsymbol{\eta}_{1,n}^{*}),g(\boldsymbol{\eta}_{2,n}^{*})), \sum_{j = 1}^2\left[\dfrac{\partial h}{\partial\theta_j}\right]^2_{\theta_j = g(\boldsymbol{\eta}_{j,n}^{*})} \left[\dfrac{\partial g}{\partial\boldsymbol{\eta}}^T \mathcal{J}^*_j(\boldsymbol{\eta})^{-1} \dfrac{\partial g}{\partial\boldsymbol{\eta}}\right]_{\boldsymbol{\eta} = \boldsymbol{\eta}_{j,n}^{*}}\right),
	\end{equation} 
\begin{equation*}\label{eqn:norm4}
\text{where}~~~~\mathcal{J}^*_j(\boldsymbol{\eta}) = n\mathcal{I}(\boldsymbol{\eta})-\dfrac{\partial^2}{\partial\boldsymbol{\eta}^2} \log(p_j(\boldsymbol{\eta})).
	\end{equation*} 
The observed information is again replaced with the Fisher information in $\mathcal{J}^*_j(\boldsymbol{\eta})$ of (\ref{eqn:norm3}) since we do not generate samples $\boldsymbol{y}^{_{(n)}}$. Algorithm \ref{alg3} details how we map a single point $\boldsymbol{u} \in [0,1]^{2d}$ to the posterior approximation in (\ref{eqn:norm3}). 

\begin{algorithm}
\caption{Mapping Posteriors to $[0,1]^{2d}$ with a Hybrid Method}
\label{alg3}

\begin{algorithmic}[1]
\setstretch{1}
\Procedure{MapHybrid}{$f(y; \boldsymbol{\eta})$, $\boldsymbol{\eta}_{1,0}$, $\boldsymbol{\eta}_{2,0}$, $g(\cdot)$, $h(\cdot)$, $n$, $\boldsymbol{u}$, $p_1(\boldsymbol{\eta}_1)$, $p_2(\boldsymbol{\eta}_2)$}
\State Generate $\hat{\boldsymbol{\eta}}_{1,n}(\boldsymbol{u})$ and $ \hat{\boldsymbol{\eta}}_{2,n}(\boldsymbol{u})$ using Lines 2 to 6 of Algorithm \ref{alg1}.
\For{$j$ in 1:2}
    \State Obtain $\boldsymbol{\eta}_{j,n}^{*}$ as $\text{arg\,max}_{{\boldsymbol{\eta}_j}}$ of the right side of (\ref{eqn:opt_exp}) anchored at $\boldsymbol{\eta}_{j,n} = \hat{\boldsymbol{\eta}}_{j,n}(\boldsymbol{u})$.
    \EndFor
    \State Use $\boldsymbol{\eta}_{1,n}^{*}$, $\boldsymbol{\eta}_{2,n}^{*}$, and the partial derivatives of $g(\cdot)$ and $h(\cdot)$ to obtain (\ref{eqn:norm3}).
\EndProcedure

\end{algorithmic}
\end{algorithm}

 \subsection{Theoretical Properties of the Power Estimates}\label{samp:theo}

Now that we have developed three algorithms to reduce the simulation dimension in a variety of settings, we consider the theoretical properties of the resulting power estimates. We introduce general notation to define power estimates for the simulation method $\zeta$, where $\zeta$ is Algorithm \ref{alg1}, \ref{alg2}, or \ref{alg3}. We let $\mathcal{N}(\underline\theta_{r}^{_{(n)}},\underline{\tau}_{r}^{_{(n)}})$ denote the relevant normal approximation to the posterior of $\theta$ corresponding to the point $\boldsymbol{u}_r \in [0,1]^{2d}$ and sample size $n$ for $r = 1, ..., m$. This approximation is respectively (\ref{eqn:bvm_mle}) for Algorithm \ref{alg1}, (\ref{eqn:norm}) for Algorithm \ref{alg2}, and (\ref{eqn:norm3}) for Algorithm \ref{alg3}. We incorporate the sample size $n$ into this notation because the mean $\underline\theta_{r}^{_{(n)}}$ depends on the sample size of the joint limiting distribution for $\hat{\boldsymbol{\eta}}_{1, n}$ and $\hat{\boldsymbol{\eta}}_{2, n}$. The variance $\underline{\tau}_{r}^{_{(n)}}$ is also an explicit function of $n$ in (\ref{eqn:bvm_mle}) and (\ref{eqn:norm3}) and an implicit function of the sample size in (\ref{eqn:norm}).  

The estimate for the posterior probability $Pr(\theta < \delta~| ~ data)$ is then
 \begin{equation}\label{eqn:power_ROPE}
p^{\delta}_{n, \boldsymbol{u}_r, \zeta} = \Phi\left(\dfrac{\delta - \underline\theta_{r}^{_{(n)}}}{\sqrt{\underline{\tau}_{r}^{_{(n)}}}}\right),
	\end{equation} 
 where $\Phi(\cdot)$ is the the CDF of the standard normal distribution. The estimates from (\ref{eqn:power_ROPE}) comprise sampling distributions of posterior probabilities after mapping to $[0,1]^{2d}$ given design values drawn from $p_D(\boldsymbol{\eta})$. These distributions should accurately approximate the exact sampling distribution of posterior probabilities; Theorem \ref{thm1} demonstrates that the exact sampling distribution of posterior probabilities induced by  (\ref{eqn:bvm_mle}) and (\ref{eqn:norm}) with data $\boldsymbol{Y}^{_{(n)}}$ converges to the sampling distribution prompted by Algorithm \ref{alg1} with pseudorandom sequences as $n \rightarrow \infty$.

      \begin{theorem}\label{thm1}
Let $\boldsymbol{\eta}_0 = (\boldsymbol{\eta}_{1,0}, \boldsymbol{\eta}_{2,0}) \sim p_D(\boldsymbol{\eta})$ for some design prior $p_D(\boldsymbol{\eta})$ such that the following conditions hold for all $\boldsymbol{\eta}_{0}$ with $p_D(\boldsymbol{\eta}_{0}) > 0$. Let $f(y; \boldsymbol{\eta}_{1,0})$ and $f(y; \boldsymbol{\eta}_{2,0})$ satisfy the regularity conditions from Appendix A.2. Let the prior $p_j(\boldsymbol{\eta}_j)$ be continuous in a neighbourhood of $\boldsymbol{\eta}_{j,0}$ with positive density at $\boldsymbol{\eta}_{j,0}$ for $j = 1, 2$. Let $g(\boldsymbol{\eta})$ and $h(\theta_1, \theta_2)$ be respectively differentiable at $\boldsymbol{\eta}_{j,0}$ and $\theta_{j,0} = g(\boldsymbol{\eta}_{j,0})$ for $j = 1,2$ with nonzero derivatives. Let $\boldsymbol{U} \overset{\text{i.i.d.}}{\sim} \mathcal{U}([0,1]^{2d})$ and $\boldsymbol{Y}^{_{(n)}}$ be generated independently from $f(y; \boldsymbol{\eta}_{1,0})$ and $f(y; \boldsymbol{\eta}_{2,0})$. Let $\mathcal{P}^{\delta}_{n, \Pi, \zeta}$ denote the sampling distribution of posterior probabilities for $Pr(\theta < \delta~| ~ data)$ given sample size $n$ produced using input $\Pi$ with method $\zeta$. Then,
\vspace*{-8pt}
 \begin{enumerate}
     \item[(a)] $\mathcal{P}^{\delta}_{n, \boldsymbol{Y}^{_{(n)}}, (\ref{eqn:bvm_mle})} \xrightarrow{d} \mathcal{P}^{\delta}_{n, \boldsymbol{U}, Alg. \ref{alg1}}$.
     \item[(b)] $\mathcal{P}^{\delta}_{n, \boldsymbol{Y}^{_{(n)}}, (\ref{eqn:norm})} \xrightarrow{d} \mathcal{P}^{\delta}_{n, \boldsymbol{U}, Alg. \ref{alg1}}$.
 \end{enumerate}
	
\end{theorem}

The proof of Theorem \ref{thm1} is given in Appendix A.3 of the supplement. While the approximations in (\ref{eqn:norm}) and (\ref{eqn:norm3}) should better account for the prior distributions for moderate sample sizes $n$, they do not differ from the approximation in (\ref{eqn:bvm_mle}) in the limit of infinite data. Likewise, $\mathcal{P}^{\delta}_{n, \boldsymbol{U}, Alg. \ref{alg2}}$ and $\mathcal{P}^{\delta}_{n, \boldsymbol{U}, Alg. \ref{alg3}}$ will converge in distribution to $\mathcal{P}^{\delta}_{n, \boldsymbol{U}, Alg. \ref{alg1}}$ as $n \rightarrow \infty$ under the conditions for Theorem \ref{thm1}. This result is straightforward because $\tilde{\boldsymbol{\eta}}_{j,n} - \hat{\boldsymbol{\eta}}_{j,n}$ and ${\boldsymbol{\eta}}_{j,n}^* - \hat{\boldsymbol{\eta}}_{j,n}$ converge in probability to 0, and $\mathcal{J}_j(\tilde{\boldsymbol{\eta}}_{j,n})/n$ and $\mathcal{J}_j^*(\boldsymbol{\eta}_{j,n}^*)/n$ converge in probability to $\mathcal{I}(\boldsymbol{\eta}_{j,0})$ given results from Appendix A.3. These results prompt Corollary \ref{cor1}, which follows from Theorem \ref{thm1}.

\begin{corollary}\label{cor1}
    Let $p^{\delta}_{n, \boldsymbol{u}_r, \zeta}$ from (\ref{eqn:power_ROPE}) be the estimate for $Pr(\theta < \delta~| ~ data)$ corresponding to sample size $n$, method $\zeta$, and point $\boldsymbol{u}_r \in [0,1]^{2d}$. Under the conditions for Theorem \ref{thm1} when $p_D(H_1) = 1$, power in (\ref{eq:power_prob}) and (\ref{eq:power_ci}) is consistently estimated by
     $$\dfrac{1}{m}\sum_{r = 1}^m\mathbb{I}\{p^{\delta_U}_{n, \boldsymbol{u}_r, \zeta} - p^{\delta_L}_{n, \boldsymbol{u}_r, \zeta} \ge \gamma\} ~~~\text{and}~~~ \dfrac{1}{m}\sum_{r = 1}^m\mathbb{I}\{p^{\delta_L}_{n, \boldsymbol{u}_r, \zeta} < \alpha/2 ~\cap~ 1 - p^{\delta_U}_{n, \boldsymbol{u}_r, \zeta} < \alpha/2\},$$
    respectively, when $\zeta$ is Algorithm \ref{alg1}, \ref{alg2}, or \ref{alg3} and $\boldsymbol{U}_r \overset{\text{i.i.d.}}{\sim} \mathcal{U}([0,1]^{2d})$ for $r = 1, ..., m$ as $n \rightarrow \infty$.
\end{corollary}

Because the limiting posterior of $\theta$ is normal when the conditions for the BvM theorem hold, power for hypothesis testing with HDIs defined in (\ref{eq:power_hdi}) should be approximated by power defined in (\ref{eq:power_ci}) for sufficiently large $n$. Corollary \ref{cor1} ensures that our three algorithms give rise to consistent power estimates as $n \rightarrow \infty$; however, it does not guarantee that these estimators are unbiased for finite $n$. When the assumptions for Theorem \ref{thm1} are satisfied, our power estimates are suitable for sufficiently large $n$. To optimize the performance of these design methods for moderate $n$, one should consider transformations of $\theta$ or certain components in $\boldsymbol{\eta}_1$ and $\boldsymbol{\eta}_2$ to improve the normal approximations to the relevant posterior and MLE distributions.

  \section{A Fast Method for Power Curve Approximation}\label{sec:curve}  

\subsection{Power Estimates with Fewer Posteriors}\label{samp:quasi}

This section details how we leverage the mappings between posterior probabilities and $[0,1]^{2d}$ to expedite power curve approximation for posterior analyses. The novelty of this computational efficiency stems from using sampling distribution segments to approximate the power curve, and our approach for power curve approximation described in Algorithm \ref{alg4} is the main contribution of this paper. Before introducing that approach, we justify why low-discrepancy sequences can be used instead of pseudorandom ones to reduce the number of posteriors required for precise power estimates. 

Unlike pseudorandom sequences, low-discrepancy sequences induce negative dependence between the points $\boldsymbol{U}_1, ..., \boldsymbol{U}_m$. We use a particular class of low-discrepancy sequences called Sobol' sequences \citep{sobol1967distribution}. Sobol' sequences are created deterministically based on integer expansion in base 2, and they are regularly incorporated into quasi-Monte Carlo methods \citep{lemieux2009using}. Randomized Sobol' sequences yield estimators with better consistency properties than those created using purely deterministic sequences, and this randomization can be carried out via a digital shift \citep{lemieux2009using}.  We generate and randomize Sobol' sequences in R using the \texttt{qrng} package \citep{hofert2020package}. When using randomized Sobol’ sequences, each point in the sequence is such that $\boldsymbol{U}_r \sim U\left([0,1]^{2d}\right)$ for $r = 1, ..., m$. It follows that randomized Sobol' sequences can be used similarly to pseudorandom sequences in Monte Carlo simulation to prompt unbiased estimators:
\begin{equation}\label{eqn:qmc_exp}
    \mathbb{E}\left(\dfrac{1}{m}\sum_{r=1}^m\Psi(\boldsymbol{U}_r)\right) = \int_{[0,1]^{2d}}\Psi(\boldsymbol{u})d\boldsymbol{u},
\end{equation}
where either indicator function in Corollary \ref{cor1} is the relevant function $\Psi(\cdot)$. Corollary \ref{cor2} formalizes this result.
\begin{corollary}\label{cor2}
    Under the conditions for Theorem \ref{thm1}, using a randomized low-discrepancy sequence $\boldsymbol{U}_1, ..., \boldsymbol{U}_m$ in lieu of a sequence $\boldsymbol{U}_r \overset{\text{i.i.d.}}{\sim} \mathcal{U}([0,1]^{2d})$ for $r = 1, ..., m$ does not impact the consistency of the power estimates from Corollary \ref{cor1} as $n \rightarrow \infty$.
\end{corollary}

Due to the negative dependence between the points, the variance of the estimator in (\ref{eqn:qmc_exp}) is typically reduced by using low-discrepancy sequences. We have that
\begin{equation}\label{eqn:qmc_var}
    \text{Var}\left(\dfrac{1}{m}\sum_{r=1}^m\Psi(\boldsymbol{U}_r)\right) = \dfrac{\text{Var}\left({\Psi(\boldsymbol{U}_r)}\right)}{m} + \dfrac{2}{m^2}\sum_{r = 1}^m\sum_{t = r + 1}^m
    \text{Cov}\left(\Psi(\boldsymbol{U}_r), \Psi(\boldsymbol{U}_t)\right),
\end{equation}
where the the first term on the right side of (\ref{eqn:qmc_var}) is the variance of the corresponding estimator based on pseudorandom sequences of independently generated points. While underutilized in experimental design, low-discrepancy sequences give rise to effective variance reduction methods when the dimension of the simulation is moderate. However, high-dimensional low-discrepancy sequences may have poor low-dimensional projections, which can lead to a deterioration in performance. This is why we proposed the low-dimensional posterior approximation methods in Section \ref{sec:sampling}. We can readily find a suitable Sobol' sequence of dimension $2d$, but it may be challenging to find an appropriate sequence of dimension $2n$ when the sample size is large.  By (\ref{eqn:qmc_var}), randomized Sobol' sequences reduce the number of posteriors for $\theta$ that we must approximate to obtain precise, consistent power estimates. 

 \subsection{Selection of Sampling Distribution Segments}\label{sec:curve.alg}
 
For a given sample size $n$, we \emph{could} obtain power estimates using the formulas in Corollary \ref{cor1} with randomized Sobol' sequences. Sample size determination could be conducted by repeating this process for various values of $n$ until a suitable sample size is found. However, such a process would waste computational resources thoroughly exploring sampling distributions for unsuitable sample sizes in order to obtain an appropriate one. We argue that consistent power estimates for a given sample size $n$ can often be obtained with only a subset of the points $\boldsymbol{u}_r \in [0,1]^{2d+1}, ~r = 1, ..., m$. As described later in Algorithm \ref{alg4}, we add a dimension to the hypercube so that we can use the final coordinate to sample $\boldsymbol{\eta}_{0}$ from a nondegenerate design prior $p_D(\boldsymbol{\eta})$. By using only a subset of points to explore most sample sizes, we consider only segments of the relevant sampling distributions.

In each of our algorithms, the approximately normal posterior $\mathcal{N}(\underline\theta_{r}^{_{(n)}},\underline{\tau}_{r}^{_{(n)}})$ and corresponding posterior probabilities $p^{\delta}_{n, \boldsymbol{u}_r, \zeta}$ depend on the design distributions, the sample size $n$, and the Sobol' sequence point $\boldsymbol{u}_r, ~ r = 1, ..., m$. Standard practice fixes the sample size $n$ and allows the point $\boldsymbol{u}_r \in [0,1]^{2d+1}$ to vary when estimating power. We now fix the point $\boldsymbol{u}_r$ and let the sample size $n$ vary. When the point $\boldsymbol{u}_r$ and design distributions are fixed, $p^{\delta}_{n, \boldsymbol{u}_r, \zeta}$ is a deterministic function of $n$. Lemma \ref{lem1} motivates our approach to choose subsets of points $\boldsymbol{u}_r \in [0,1]^{2d+1}$ for each sample size $n$ explored. 
\begin{lemma}\label{lem1}
    Let the conditions for Theorem \ref{thm1} be satisfied. For a given point $\boldsymbol{u}_r = (u_1, ..., u_{2d+1}) \in [0,1]^{2d+1}$, we have that Algorithms \ref{alg1}, \ref{alg2}, and \ref{alg3} prompt  
    \vspace*{0pt}
 \begin{enumerate}
     \item[(a)] $p^{\delta}_{n, \boldsymbol{u}_r, \zeta} \approx \Phi\left(a(\delta, \theta_0)\sqrt{n} + b(\boldsymbol{u}_{r})\right)$ for sufficiently large $n$, where $\theta_0 = h(g(\boldsymbol{\eta}_{1,0}, \boldsymbol{\eta}_{2,0}))$ and $a(\cdot)$ and $b(\cdot)$ are functions that do not depend on $n$.
     \item[(b)] When $\theta_0 \in (\delta_L, \delta_U)$, $p^{\delta_U}_{n, \boldsymbol{u}_r, \zeta} - p^{\delta_L}_{n, \boldsymbol{u}_r, \zeta}$ is a increasing function of $n$ for sufficiently large sample sizes.
 \end{enumerate}
\end{lemma} 
We prove Lemma \ref{lem1} in Appendix B of the supplement and now consider its implications when $\theta_0 \in (\delta_L, \delta_U)$ for a given point $\boldsymbol{u}_r$. Design priors $p_D(\boldsymbol{\eta})$ that prompt $p_D(H_1) = 1$ ensure $\theta_0 \in (\delta_L, \delta_U)$. If $p^{\delta_U}_{n_A, \boldsymbol{u}_r, \zeta} - p^{\delta_L}_{n_A, \boldsymbol{u}_r, \zeta} \ge \gamma$, then $p^{\delta_U}_{n_B, \boldsymbol{u}_r, \zeta} - p^{\delta_L}_{n_B, \boldsymbol{u}_r, \zeta} \ge \gamma$ for sufficiently large $n_A < n_B$. The $(\theta_1, \theta_2)$-space such that $\theta = h(\theta_1, \theta_2) \in (\delta_L, \delta_U)$ is convex when $\theta = \theta_1 - \theta_2$ or $\theta = \theta_1 /\theta_2$. This convexity limits the potential for decreasing behaviour of $p^{\delta_U}_{n, \boldsymbol{u}_r, \zeta} - p^{\delta_L}_{n, \boldsymbol{u}_r, \zeta}$ as a function of $n$ for small and moderate sample sizes. 

In light of this, our method to approximate the power curve generates a single Sobol' sequence of length $m$. For hypothesis tests with posterior probabilities, we use root finding algorithms \citep{brent1973algorithm} to find the value for $n$ such that $p^{\delta_U}_{n, \boldsymbol{u}_r, \zeta} - p^{\delta_L}_{n, \boldsymbol{u}_r, \zeta} - \gamma = 0$ for $r = 1, ..., m$. The empirical CDF of these $m$ sample sizes approximates the power curve. As detailed in Section \ref{sec:curve.target}, this root-finding approach facilitates targeted exploration of $[0,1]^{2d+1}$ based on the sample size $n$. Since the posterior probabilities in Corollary \ref{cor1} are mapped to $\boldsymbol{u}_r \in [0,1]^{2d+1}$, this approach also allows us to consider segments of the approximate sampling distribution of posterior probabilities. 

  \subsection{Power Curves with Sampling Distribution Segments}\label{sec:curve.target}

Algorithm \ref{alg4} formally describes our method for power curve approximation with sampling distribution segments for analyses facilitated via posterior probabilities and Bayes factors. We later discuss the necessary modifications for analyses with credible intervals. To implement this approach, we must choose a parametric statistical model $f(y; \boldsymbol{\eta})$, functions $g(\cdot)$ and $h(\cdot)$, analysis priors $p_1(\boldsymbol{\eta}_1)$ and $p_2(\boldsymbol{\eta}_2)$, and a design prior that prompts $({\boldsymbol{\eta}}_{1,0}, {\boldsymbol{\eta}}_{2,0}) \sim p_D(\boldsymbol{\eta})$. We recommend using visualization techniques and consulting the literature on prior elicitation to choose $p_D(\boldsymbol{\eta})$ \citep{chaloner1996elicitation,garthwaite2005statistical,johnson2010methods}. We must also select an interval $(\delta_L, \delta_U)$, a critical value $\gamma$, a target power $\Gamma$, a method $\zeta$ consisting of one of the three algorithms proposed in Section \ref{sec:sampling}, and the length of the Sobol' sequence $m$. We elaborate on how to choose $m$ and the design prior $p_D(\boldsymbol{\eta})$ as part of our numerical studies in Section \ref{sec:study}.

\begin{algorithm}
\caption{Procedure for Power Curve Approximation with Posterior Probabilities}
\label{alg4}

\begin{algorithmic}[1]
\setstretch{1}
\Procedure{PowerCurve}{$f(y; \boldsymbol{\eta})$, $g(\cdot)$, $h(\cdot)$, $p_j(\boldsymbol{\eta}_j)$, $p_D(\boldsymbol{\eta})$, $(\delta_L, \delta_U)$, $\gamma$, $\Gamma$, $\zeta$, $m$}
\State Generate $m$ draws $\boldsymbol{\eta}_{0[1]}, \dots, \boldsymbol{\eta}_{0[m]}$ from $p_D(\boldsymbol{\eta})$.
\State Generate Sobol' sequence $\{\boldsymbol{u}_r\}_{r = 1}^m \in [0,1]^{2d+1}$.
\State Reorder $\{\boldsymbol{\eta}_{0[r]}\}_{r = 1}^m$ so $\theta_{0 [r]} = h(g(\boldsymbol{\eta}_{1, 0[r]}, \boldsymbol{\eta}_{2, 0[r]}))$ prompts the $\left\lceil m \boldsymbol{u}_{r [2d + 1]} \right\rceil^{\text{th}}$ smallest realization of  \linebreak \hspace*{12pt} $\{\theta_{0 [r]}\}_{r = 1}^m$.
\State Let $n_0$ equate (\ref{eq:power_prob}) to $\Gamma$ when $p(\boldsymbol{\eta}_j~|~ data) \approx \mathcal{N}(\hat{\boldsymbol{\eta}}_{j,n}, \mathcal{I}(\boldsymbol{\eta}_{j,0})^{-1}/n)$, where  $\hat{\boldsymbol{\eta}}_{j,n} \sim \mathcal{N}({\boldsymbol{\eta}}_{j,0}, \mathcal{I}(\boldsymbol{\eta}_{j,0})^{-1}/n)$ \linebreak \hspace*{12pt} and $(\boldsymbol{\eta}_{1,0}, \boldsymbol{\eta}_{2,0}) \sim p_D(\boldsymbol{\eta})$.
 \State \texttt{sampSobol} $\leftarrow \textsc{null}$
 \For{$r$ in 1:$m$}
    \State Let $\texttt{sampSobol}_{[r]}$ solve $p^{\delta_U}_{n, \boldsymbol{u}_r, \zeta} - p^{\delta_L}_{n, \boldsymbol{u}_r, \zeta} - \gamma = 0$ in terms of $n$ for $\boldsymbol{\eta}_0 = \boldsymbol{\eta}_{0[r]}$,  initializing the \linebreak \hspace*{28pt} root-finding algorithm at $\lceil n_0 \rceil$.
 \EndFor
    \State $\texttt{powerCurve} \leftarrow$ empirical CDF of $\texttt{sampSobol}$
    \State Let $n_*$ be the $\Gamma$-quantile of $\texttt{sampSobol}$.
    \For{$r$ in 1:$m$}
 \If{$\texttt{sampSobol}_{[r]} \le n_*$}
 \If{$p^{\delta_U}_{n_*, \boldsymbol{u}_r, \zeta} - p^{\delta_L}_{n_*, \boldsymbol{u}_r, \zeta} - \gamma < 0$}
 \State Repeat Line 8, initializing the root-finding algorithm at $n_*$.
 \EndIf
    \Else
    \If{$p^{\delta_U}_{n_*, \boldsymbol{u}_r, \zeta} - p^{\delta_L}_{n_*, \boldsymbol{u}_r, \zeta} - \gamma \ge 0$}
 \State Repeat Line 8, initializing the root-finding algorithm at $n_*$.
 \EndIf
    \EndIf
 \EndFor
 \State $\texttt{powerCurveFinal} \leftarrow$ empirical CDF of $\texttt{sampSobol}$
    \State Let $n^*$ be the $\Gamma$-quantile of $\texttt{sampSobol}$.
 \State \Return $\texttt{powerCurveFinal}$, $\lceil n^* \rceil$ as recommended sample size
\EndProcedure

\end{algorithmic}
\end{algorithm}

 Lines 2 to 4 of Algorithm \ref{alg4} initialize our procedure for the predictive approach where $p_D(\boldsymbol{\eta})$ is nondegenerate. After generating $m$ design values $\boldsymbol{\eta}_0 = (\boldsymbol{\eta}_{1,0}, \boldsymbol{\eta}_{2,0})$ from $p_D(\boldsymbol{\eta})$ and $m$ Sobol' sequence points from $[0,1]^{2d+1}$, we reorder the draws $\{\boldsymbol{\eta}_{0[r]}\}_{r = 1}^m$ in Line 4. We use the final coordinate of each point $\boldsymbol{u}_{r}$ to reorder those draws with respect to their design value for $\theta_0 = h(g(\boldsymbol{\eta}_{1,0}), g(\boldsymbol{\eta}_{2,0}))$. This reordering allows us to join the design values and maximum likelihood estimates prompted by the method $\zeta$ and the first $2d$ coordinates of $\boldsymbol{u}_{r}$ in a low-discrepancy manner. If we use the conditional approach where $p_D(\boldsymbol{\eta})$ is degenerate, all $m$ design values $\boldsymbol{\eta}_0$ in Line 2 are identical. In that case, we can generate a Sobol' sequence of dimension $2d$ in Line 3 and skip Line 4. 


 Line 5 of Algorithm \ref{alg4} uses the normal approximation in (\ref{eqn:bvm_mle}) with known variance to obtain a starting point $n_0$ for the root-finding algorithm. We can obtain this starting point in a fraction of a second, and it should be close to the final sample size recommendation if uninformative priors are used. Since the root-finding algorithm is initialized at $\lceil n_0 \rceil$ for all points $\boldsymbol{u}_r$, $r = 1, ..., m$, the entire sampling distribution of posterior probabilities is explored for that sample size. In Lines 7 and 8 of Algorithm \ref{alg4}, the root-finding algorithm then facilitates segmented exploration of the approximate distribution of posterior probabilities for all other sample sizes considered. We complete the power curve approximation procedure by exploring the entire sampling distribution of posterior probabilities at the sample size $n_*$ in Lines 11 to 17. If the statements in Lines 13 or 16 are true, this implies that $p^{\delta_U}_{n, \boldsymbol{u}_r, \zeta} - p^{\delta_L}_{n, \boldsymbol{u}_r, \zeta} = \gamma$ for at least two distinct sample sizes $n$. For these points $\boldsymbol{u}_r$, we can reinitialize the root-finding algorithm at $n_*$ to obtain a solution for each point that will make the power curve consistent at $n_*$. This consistency is a direct consequence of Corollary \ref{cor2}.

 In Section \ref{sec:study}, we conduct numerical studies to demonstrate the suitability of the power curves obtained by Algorithm \ref{alg4} in various settings. These numerical results show that the if statements in Lines 13 and 16 are very rarely true for any point $\boldsymbol{u}_r \in [0,1]^{2d+1}$ when sample sizes are large enough for the BvM theorem to hold. In those situations, $n_* = n^*$ and both the power estimate at $n^*$ and the sample size recommendation $\lceil n^* \rceil$ are consistent. It is incredibly unlikely that $n_*$ and $n^*$ would differ substantially, but Lines 11 to 17 of Algorithm \ref{alg4} could be repeated in that event, where the root-finding algorithm is initialized at $n^*$ instead of $n_*$. These consistent sample size recommendations are not guaranteed to be unbiased since the normal approximation to the relevant posterior and MLE distributions may introduce noticeable bias for finite $n$. Even though the power curves from Algorithm \ref{alg4} are consistent near the target power $\Gamma$, their global consistency at all sample sizes $n$ is not guaranteed because Lemma \ref{lem1} and the BvM theorem are large-sample results. Nevertheless, our numerical studies in Section \ref{sec:study} highlight good global estimation of the power curve, particularly when the method $\zeta$ for posterior mapping accounts for the priors. 

 We discuss how to generalize Algorithm \ref{alg4} for analyses with credible intervals below. If one of $\delta_L$ or $\delta_U$ is not finite, Algorithm \ref{alg4} can be used without modification where $\gamma = 1 - \alpha$. Otherwise, the initial value for the root-finding algorithm $n_0$ found in Line 5 is based on (\ref{eq:power_ci}) instead of (\ref{eq:power_prob}). In Line 8, $\texttt{sampSobol}_{[r]}$ is modified to be the maximum of the two solutions for $1 - p^{\delta_L}_{n, \boldsymbol{u}_r, \zeta} - (1-\alpha/2) = 0$ and $p^{\delta_U}_{n, \boldsymbol{u}_r, \zeta} - (1-\alpha/2) = 0$. We lastly modify how the sampling distribution of posterior probabilities at the sample size $n_*$ is explored. To do so, we confirm that both $1-p^{\delta_L}_{n_*, \boldsymbol{u}_r, \zeta} - (1-\alpha/2)$ and $p^{\delta_U}_{n_*, \boldsymbol{u}_r, \zeta} - (1-\alpha/2)$ are not less than or at least 0 in Lines 13 and 16, respectively. Effectively, power curve approximation with equal-tailed credible intervals requires us to consider two hypothesis tests with posterior probabilities and intervals $(\delta_L, \infty)$ and $(-\infty, \delta_U)$. 
 
 The root-finding algorithm from Lines 7 and 8 of Algorithm \ref{alg4} approximates posteriors corresponding to $\mathcal{O}(\log_2B)$ points from $[0,1]^{2d+1}$, where $B$ is the maximum sample size considered for the power curve. We would require $\mathcal{O}(B)$ such points to explore a similar range of sample sizes using power estimates obtained via Corollary \ref{cor1} with randomized Sobol' sequences. When $B \ge 59$, our approach prompts at least an order of magnitude reduction in the number of posterior approximations because $O(\text{log}_2B) < O(B)/10$. Moreover, using randomized Sobol' sequences in lieu of pseudorandom ones has generally allowed us to estimate power curves with similar precision using about an order of magnitude fewer points from $[0,1]^{2d+1}$. We visualize the insights from this paragraph using a simpler example than the one with gamma tail probabilities in Appendix C of the supplement. 

	
	\section{Numerical Studies}\label{sec:study}
	
	\subsection{Power Curve Approximation with the Gamma Distribution}\label{sec:sim.3}

      We now compare the performance of our power curve approximation procedure across several scenarios. For each scenario, we specify design priors for the gamma tail probability example from Section \ref{sec:ex.1}. We first specify a degenerate design prior $p_D(\boldsymbol{\eta})$ under the conditional approach. To do so, we just need to choose a set of design values $\boldsymbol{\eta}_{1,0}$ and $\boldsymbol{\eta}_{2,0}$ such that $H_1$ is true. Because the ENIGH survey is conducted biennially, we choose these design values for both gamma distributions using data from the ENIGH 2018 survey \citep{enigh2018}. We repeat the process detailed in Section \ref{sec:ex.1} to create a similar data set of 2018 quarterly food expenditure per person. We adjust each expenditure to account for inflation, compounding 2\% annually, between 2018 and 2020. We find the posterior means for the gamma shape and rate parameters to be $\bar\alpha_1 = 2.11$ and $\bar\beta_1 = 0.69$ for the female provider group and $\bar\alpha_2 = 2.43$ and $\bar\beta_2 = 0.79$ for the male provider group. These posterior means comprise the design values $\boldsymbol{\eta}_{1,0}$ and $\boldsymbol{\eta}_{2,0}$. After accounting for inflation, the 2018 estimate for the median quarterly food expenditure per person in upper income households is 4.29 (MXN \$1000). For the purposes of sample size determination, we use $\kappa_0 = 4.29$ as the threshold for the gamma tail probabilities.
      
      The scenarios we consider are based on two sets of analysis priors. For the first set, we specify uninformative $\text{GAMMA}(2, 0.25)$ priors for the gamma parameters $\alpha_j$ and $\beta_j$ for group $j = 1, 2$. To choose the second set of priors, we reconsider the approximately gamma distributed posteriors used to obtain design values for $\alpha_1$, $\beta_1$, $\alpha_2$, and $\beta_2$. To incorporate prior information, we consider gamma distributions that have the same modes with variances that are larger by a factor of 10. In comparison to the $\text{GAMMA}(2, 0.25)$ prior, these distributions are quite informative. These distributions -- which we use as the set of informative priors -- are $\text{GAMMA}(34.23,  15.85)$ for $\alpha_1$, $\text{GAMMA}(27.20,  38.15)$ for $\beta_1$, $\text{GAMMA}(105.31,  42.96)$ for $\alpha_2$, and $\text{GAMMA}(85.49, 106.58)$ for $\beta_2$.
      
      For each prior specification, we first consider the quality of power curve estimation for analyses with posterior probabilities. These posterior probabilities reflect a ratio-based comparison of $\theta_1/\theta_2$, where $\theta_j$ is the gamma tail probability for group $j$ described in Section \ref{sec:ex.1}. We consider three $((\delta_L, \delta_U),\gamma, \Gamma)$ combinations: $\{a, b, c\} = \{((1.25^{-1}, 1.25), 0.5, 0.6)$, $((1.3^{-1}, \infty), 0.9, 0.7)$, $((1.15^{-1}, 1.15), 0.8, 0.8)\}$. The first combination explores moderate sample sizes. The second combination considers a one-sided noninferiority hypothesis for $\theta_1$. The third combination explores larger sample sizes. All three combinations are such that $\theta_0 = h(g(\boldsymbol{\eta}_{1,0}), g(\boldsymbol{\eta}_{2,0})) \in (\delta_L, \delta_U)$. This gives rise to six settings, each consisting of an analysis prior specification (Setting 1 = uninformative, Setting 2 = informative) and $((\delta_L, \delta_U),\gamma, \Gamma)$ combination. 
      
\begin{figure}[!tb] \centering 
		\includegraphics[width = 0.95\textwidth]{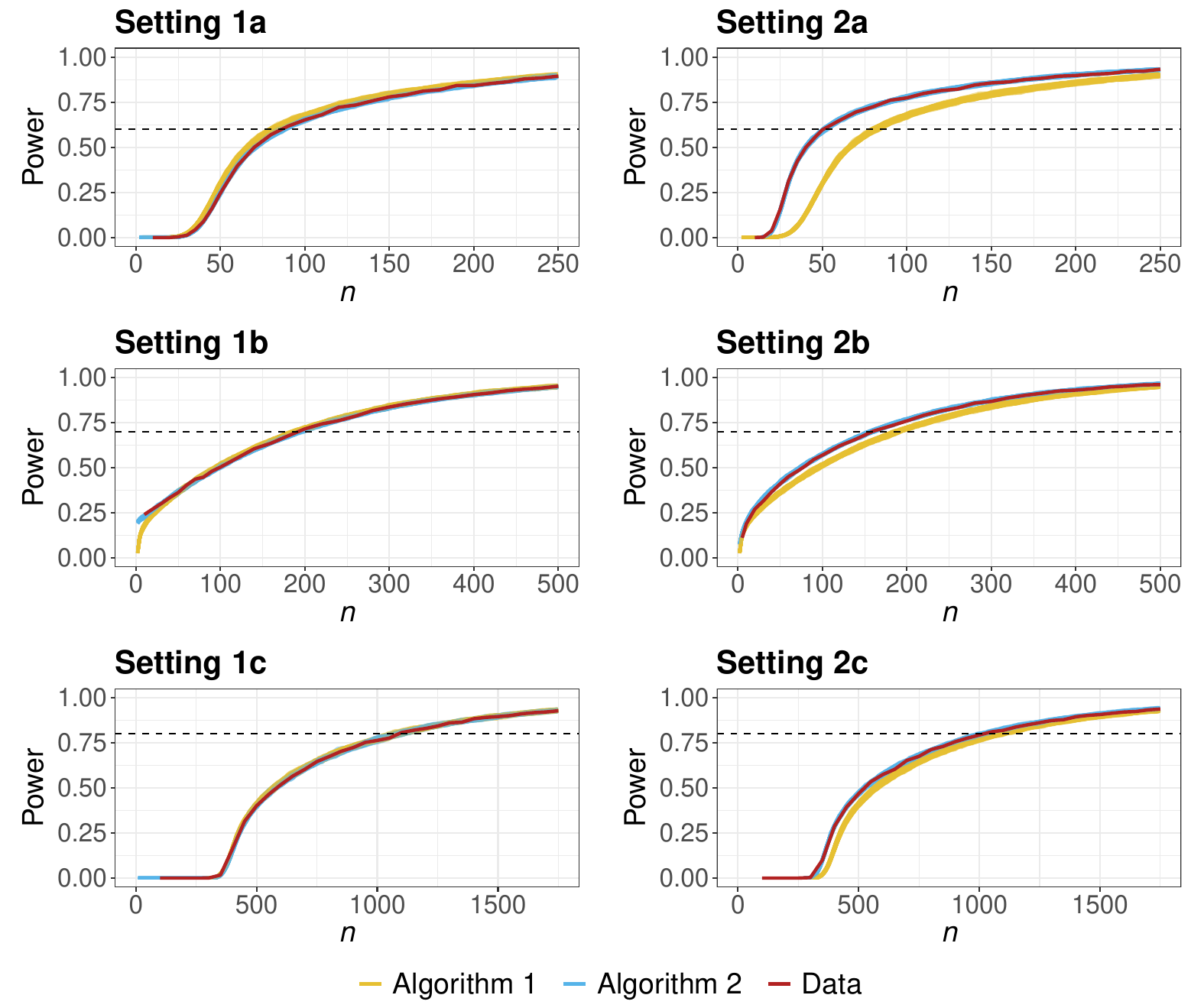} 
		\caption{\label{fig:power1} 100 power curves obtained via Algorithms \ref{alg1} (yellow) and \ref{alg2} (blue), power curve estimated via simulated data (red), and target power $\Gamma$ (dotted line) for each setting with hypothesis tests facilitated via posterior probabilities.} 
	\end{figure}

      For each setting, we generated 100 power curves using Algorithm \ref{alg4} with $\zeta = \{\text{Alg. \ref{alg1}}, \text{Alg. \ref{alg2}}\}$. We do not consider Algorithm \ref{alg3} for this example because the gamma model belongs to the exponential family. The conditional approach is employed to estimate these power curves, so we generate a Sobol' sequence $\{\boldsymbol{u}_r \}_{r = 1}^m \in [0,1]^{2d}$ with $d =2$ for each curve. We use $m = 1024$ with the conditional approach to balance the computational efficiency and precision of the approximation to the power curve. We also used the following transformations to improve the quality of the normal approximations for moderate $n$: $\theta = \log(\theta_1) - \log(\theta_2)$ and $\boldsymbol{\eta}_j = (\log(\alpha_j), \log(\beta_j))$ for $j = 1, 2$. We then selected an appropriate array of sample sizes $n$. For each value of $n$, we generated $10^{4}$ samples of that size from $f(y; \boldsymbol{\eta}_{1,0})$ and $f(y; \boldsymbol{\eta}_{2,0})$. We approximated the corresponding posterior of $\theta = \theta_1 /\theta_2$ using MCMC methods and determined whether $100\times\gamma\%$ of the posterior was contained within $(\delta_L, \delta_U)$. For each $n$ explored, we computed the proportion of the $10^{4}$ samples in which this occurred to approximate the power curve based on entire sampling distributions. Figure \ref{fig:power1} depicts these results.

 For all settings considered, the alignment between the blue and red curves in Figure \ref{fig:power1} indicates that the power curves generated by our method with Algorithm \ref{alg2} perform well. Considering sampling distribution segments via the root-finding algorithm has not led to performance issues -- even with moderate sample sizes. In total, we approximated 3200 power curves for the gamma tail probability example in this section and Appendix D of the supplement. We did not need to reinitialize the root-finding algorithm in Lines 11 to 17 of Algorithm \ref{alg4} for \emph{any} of the $4.506 \times 10^6$ points used to generate these 3200 curves. 
 
 The power curves obtained using our method with Algorithm \ref{alg1} require large sample sizes $n$ or uninformative priors to yield good performance. This is evident as the yellow curves do not approximate the red curves for the settings with informative priors. Even in Setting 1a with uninformative priors and moderate sample sizes, the yellow power curves are noticeably shifted to the left. The bias is  a result of the approximation in (\ref{eqn:bvm_mle}) not accounting for the priors and not the root-finding algorithm.  Algorithm \ref{alg2} also performs better for smaller sample sizes in Setting 1b as the sample sizes corresponding to low power are too small for the BvM theorem to be invoked. For Setting 1c, neither the blue nor yellow power curves differ substantially from the red curve. This is a direct consequence of the BvM theorem. 
 
 Each yellow power curve in Figure \ref{fig:power1} was estimated in 2 to 3 seconds without parallelization, whereas each blue curve took roughly 5 seconds to approximate without parallelization. Each red curve took between 2 and 4 hours to estimate using heavy parallelization with 72 cores. This longer runtime for the red curves even takes into account not estimating power at every sample size $n$. The blue curves take slightly longer to estimate than the yellow ones because we must find the posterior modes $\tilde{\boldsymbol{\eta}}_{1,n}$ and $\tilde{\boldsymbol{\eta}}_{2,n}$ using optimization methods. We therefore recommend using Algorithm \ref{alg4} with the method for posterior mapping from Algorithm \ref{alg2} whenever possible since this method accounts for the prior distributions without a substantial increase in runtime. With the same computational resources for this example, we can approximate a handful of posteriors using standard computational methods. This is not sufficient to produce even a crude power estimate for a single sample size $n$ on the red power curve.

\begin{figure}[!tb] \centering 
		\includegraphics[width = 0.95\textwidth]{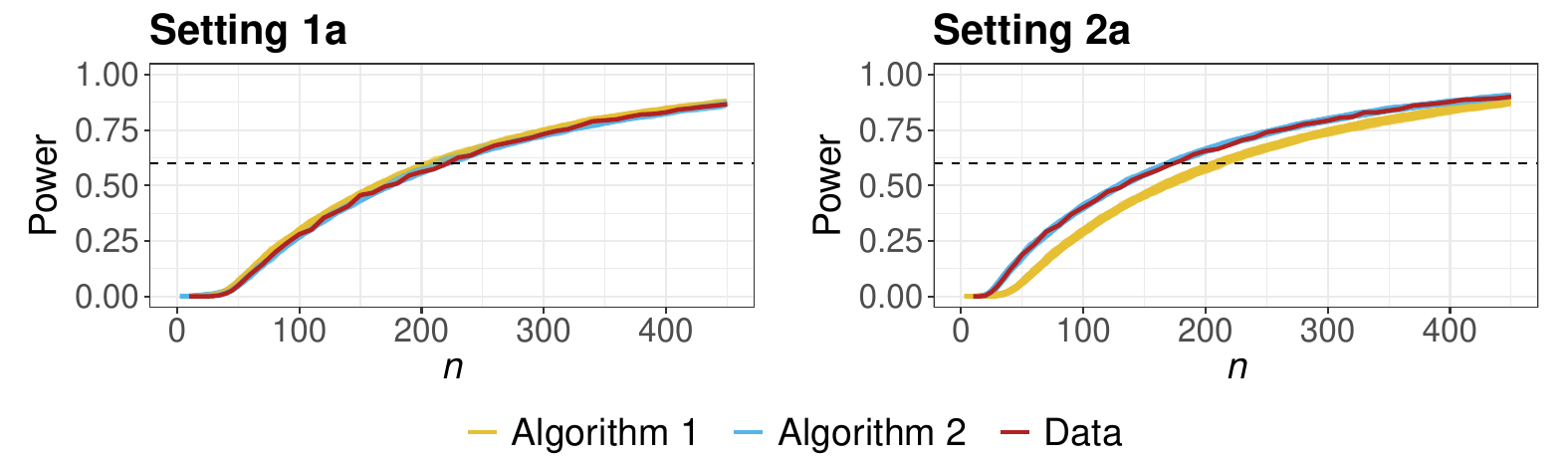} 
		\caption{\label{fig:power2} 100 power curves obtained via Algorithms \ref{alg1} (yellow) and \ref{alg2} (blue), power curve estimated via simulated data (red), and target power $\Gamma$ (dotted line) for Settings 1a and 2a with hypothesis tests facilitated via credible intervals.} 
	\end{figure}

 Because power curve approximation for hypothesis tests with Bayes factors just requires that we choose $\gamma$ to align with the right side of (\ref{eqn:bf_pp}), we consider the performance of our method for such analyses in Appendix D.1 of the supplement. We now reconsider Settings 1a and 2a with analyses facilitated via equal-tailed credible intervals under the conditional approach. We choose $\alpha = 1 - \gamma = 0.4$ for this analysis. We again implemented Algorithm \ref{alg4} with $\zeta = \{\text{Alg. \ref{alg1}}, \text{Alg. \ref{alg2}}\}$ and $m = 1024$ to obtain 100 power curves with each method. When approximating the power curve by simulating data to estimate entire sampling distributions, we computed power as the proportion of simulation repetitions in which $100\times(1 - \alpha/2)\%$ of the posterior for $\theta = \theta_1/\theta_2$ was contained within each of the following intervals: $(\delta_L, \infty)$ and $(-\infty, \delta_U)$. These results for Settings 1a and 2a are visualized in Figure \ref{fig:power2}.  For analyses with credible intervals, we draw similar conclusions about the performance of our approach with Algorithms \ref{alg1} and \ref{alg2} as in Figure \ref{fig:power1}. Each blue curve in Figure \ref{fig:power2} took roughly 7 seconds to estimate since we must examine the intervals corresponding to two one-sided tests with posterior probabilities as discussed in Section \ref{sec:curve.target}.

 We now consider the predictive approach with hypothesis tests facilitated via posterior probabilities for Settings 1a and 2a. These settings assess the hypothesis $H_1: \theta = \theta_1/\theta_2 \in (1.25^{-1}, 1.25)$. The predictive approach requires that we specify a nondegenerate design prior $p_D(\boldsymbol{\eta})$. We now reconsider the informative analysis priors for $\alpha_1$, $\beta_1$, $\alpha_2$, and $\beta_2$ introduced in the second paragraph of this subsection. When we independently join those four gamma priors, it induces a prior on $\theta$. To consider power in (\ref{eq:power_prob}), our design prior must be such that $p_D(H_1) = 1$. However, this induced prior on $\theta$ is such that $p_D(H_1) =  0.2765$. We therefore take a segment of the informative analysis priors as the design prior $p_D(\boldsymbol{\eta})$. This segment of the analysis priors on $\boldsymbol{\eta}_1 = (\alpha_1, \beta_1)$ and $\boldsymbol{\eta}_2 = (\alpha_2, \beta_2)$ is chosen to be such that $\theta = h(g(\boldsymbol{\eta}_1), g(\boldsymbol{\eta}_2)) \in (1.1^{-1}, 1.1)$ for illustrative purposes. We define $p_D(\boldsymbol{\eta})$ such that the design values $\boldsymbol{\eta}_{1,0}$ and $\boldsymbol{\eta}_{2,0}$ do not prompt a $\theta_0$ value that is near the endpoints of the interval $(1.25^{-1}, 1.25)$. If $\theta_0$ is very close to $\delta_L$ or $\delta_U$, then we may require an impractically large sample size $n$ for the posterior of $\theta$ to be concentrated in the interval $(\delta_L, \delta_U)$. We emphasize that this approach is just one possible way of defining a suitable design prior. This example primarily serves to illustrate that we can approximate power curves while incorporating uncertainty in the design values $\boldsymbol{\eta}_{1,0}$ and $\boldsymbol{\eta}_{2,0}$.

 \begin{figure}[!tb] \centering 
		\includegraphics[width = 0.95\textwidth]{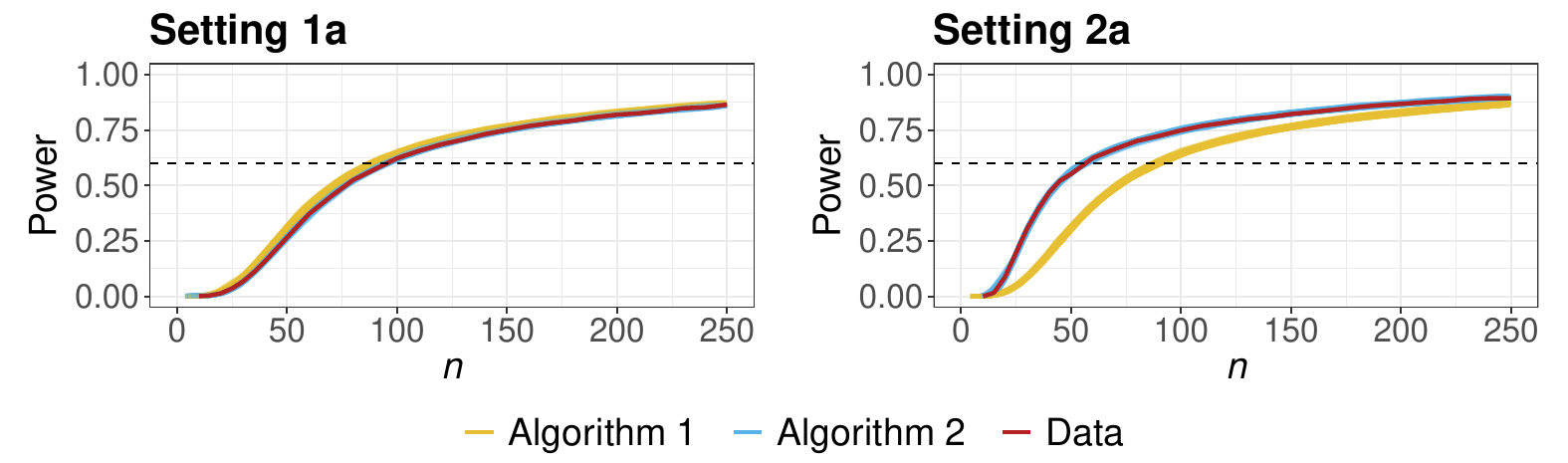} 
		\caption{\label{fig:power.pred} 100 power curves obtained via Algorithms \ref{alg1} (yellow) and \ref{alg2} (blue), power curve estimated via simulated data (red), and target power $\Gamma$ (dotted line) for Settings 1a and 2a with hypothesis tests using the predictive approach.} 
	\end{figure}

 To implement Algorithm \ref{alg4}, we must obtain a sample of $m$ draws from $p_D(\boldsymbol{\eta})$. We can obtain these draws from the design prior defined in the previous paragraph using rejection sampling \citep{casella2004generalized}. With the predictive approach, we use $m = 4096$ instead of $m = 1024$ considered for the conditional approach. We increase the number of simulation repetitions because sampling from a nondegenerate design prior introduces an additional source of variability that must be averaged over. We also require a Sobol' sequence of dimension $2d + 1 = 5$ to approximate each power curve: $\{\boldsymbol{u}_r \}_{r=1}^{4096} \in [0,1]^5$. For each setting, we generated 100 power curves using Algorithm \ref{alg4} with $\zeta = \{\text{Alg. \ref{alg1}}, \text{Alg. \ref{alg2}}\}$ and the monotonic transformations of $\theta$ and $\boldsymbol{\eta}_j$ described earlier. For each value of $n$ in an appropriate array of sample sizes, we generated $2\times10^{4}$ samples of that size from $f(y; \boldsymbol{\eta}_{1,0})$ and $f(y; \boldsymbol{\eta}_{2,0})$, where $\boldsymbol{\eta}_0 = (\boldsymbol{\eta}_{1,0}, \boldsymbol{\eta}_{2,0})$ was independently drawn from $p_D(\boldsymbol{\eta})$ in each simulation repetition. We used these data to approximate the posterior of $\theta = \theta_1/\theta_2$ and estimate power curves as done in Figure \ref{fig:power1}. Figure \ref{fig:power.pred} depicts the results from this numerical study with the predictive approach.

 For this particular design prior $p_D(\boldsymbol{\eta})$, the power curves in Figure \ref{fig:power.pred} look similar to the power curves for Settings 1a and 2a in Figure \ref{fig:power1}. We therefore draw similar conclusions about the performance of our method with Algorithms \ref{alg1} and \ref{alg2} using the predictive approach as in Figure \ref{fig:power1} with the conditional approach. Each yellow and blue curve in Figure \ref{fig:power.pred} took approximately 10 and 20 seconds to estimate, respectively. These yellow and blue curves took longer to estimate than the power curves under the conditional approach because we used $m = 4096$ instead of $m = 1024$. Each red power curve took about 5 hours to estimate using parallelization with 72 cores. Because sufficient statistics can be readily computed for this example, we did not consider the performance of our approach to power analysis with the method for posterior mapping in Algorithm \ref{alg3}. We consider the performance of that approach in Section \ref{sec:ssd.6}.
	
	
		 \subsection{Power Curve Approximation with the Weibull Distribution}\label{sec:ssd.6}
	
To further explore the performance of our power curve approximation procedure, we reconsider the food expenditure example with Weibull distributions. For concision, we only consider the conditional approach in this subsection. To do so, we find design values for the Weibull distributions using the data from the ENIGH 2018 survey processed in Section \ref{sec:sim.3}. We find the posterior means for the Weibull shape and scale parameters to be $\bar\nu_1 = 1.41$ and $\bar\lambda_1 = 3.39$ for the female provider group and $\bar\nu_2 = 1.49$ and $\bar\lambda_2 = 3.42$ for the male provider group, where $\text{GAMMA}(2, 1)$ priors were assigned to each parameter. These posterior means comprise the new design values $\boldsymbol{\eta}_{1,0}$ and $\boldsymbol{\eta}_{2,0}$. As in Section \ref{sec:sim.3}, a threshold of $\kappa_0 = 4.29$ defines the Weibull tail probabilities.

      We again consider two sets of analysis priors. For the first set, we specify uninformative $\text{GAMMA}(2, 1)$ priors for the Weibull parameters $\nu_j$ and $\lambda_j$ for group $j = 1, 2$. To choose the second set of priors, we reconsider the approximately gamma distributed posteriors used to obtain design values for $\nu_1$, $\lambda_1$, $\nu_2$, and $\lambda_2$. To incorporate prior information, we consider gamma distributions that have the same modes with variances that are larger by a factor of $100$. These distributions prompt the following informative priors: $\text{GAMMA}(12.73, 8.28)$ for $\nu_1$, $\text{GAMMA}(11.81, 3.20)$ for $\lambda_1$, $\text{GAMMA}(38.35, 25.09)$ for $\nu_2$, and $\text{GAMMA}(37.91, 10.79)$ for $\lambda_2$. 
      
\begin{figure}[!b] \centering 
		\includegraphics[width = 0.95\textwidth]{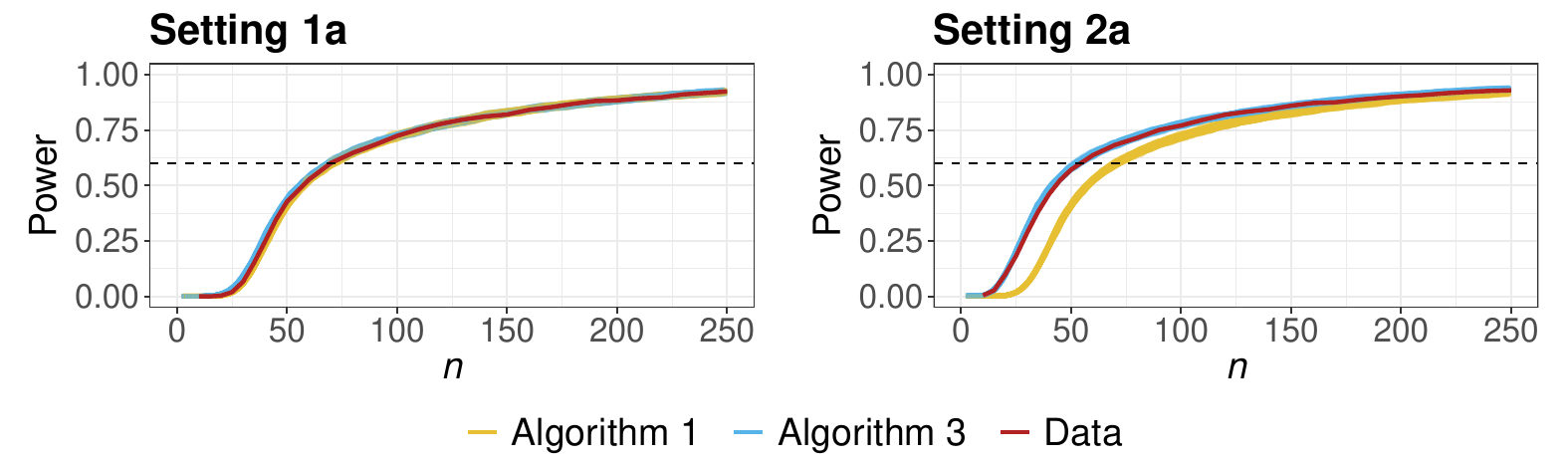} 
		\caption{\label{fig:power3} 100 power curves obtained via Algorithms \ref{alg1} (yellow) and \ref{alg3} (blue), power curve estimated via simulated data (red), and target power $\Gamma$ (dotted line) for Settings 1a and 2a with hypothesis tests using the Weibull distribution.} 
	\end{figure}

      For each analysis prior specification, we consider power curve estimation for analyses facilitated via posterior probabilities with Settings 1a and 2a from Section \ref{sec:sim.3}, where $\theta = \theta_1/\theta_2$ and $((\delta_L, \delta_U),\gamma, \Gamma) = ((1.25^{-1}, 1.25), 0.5, 0.6)$.  For each setting, we generated 100 power curves using Algorithm \ref{alg4} with $\zeta = \{\text{Alg. \ref{alg1}}, \text{Alg. \ref{alg3}}\}$ and Sobol' sequences $\{\boldsymbol{u}_r \}_{r=1}^{1024} \in [0,1]^4$. The design values for this subsection prompt $\theta_0 = 1.008 \in (\delta_L, \delta_U)$. We used the following transformations to improve the quality of the normal approximations: $\theta = \log(\theta_1) - \log(\theta_2)$ and $\boldsymbol{\eta}_j = (\log(\nu_j), \log(\lambda_j))$ for $j = 1, 2$. As in Section \ref{sec:sim.3}, we also estimated the power curves by generating data from the design distributions and approximating the posterior of $\theta = \theta_1/\theta_2$. Figure \ref{fig:power3} depicts these results. 

   For Settings 1a and 2a, we draw similar conclusions for the blue and yellow power curves as in Figure \ref{fig:power1} with the gamma model. Algorithm \ref{alg3} therefore yields suitable performance for this example when low-dimensional sufficient statistics cannot be calculated. Each yellow and blue power curve in Figure \ref{fig:power3} was estimated in less than five seconds without parallelization. For this example with the Weibull model, we cannot approximate a single posterior of $\theta$ using the same computing resources. Each red power curve took over 12 hours to estimate using parallelization with 72 cores. While our power curve approximation method does not require parallel computing for fast performance, all yellow and blue power curves in our numerical studies with the conditional approach could be estimated in a second or two if parallelized on a standard laptop with four cores. All yellow and blue power curves with the predictive approach could be estimated in up to five seconds with the same computing resources. As such, our method for power curve approximation allows users to quickly explore potential designs for their study in real time, expediting communication between stakeholders of the study.

	
	\section{Discussion}\label{sec:conclusion}
	
	In this paper, we developed a framework for fast power curve approximation with hypothesis tests facilitated via posterior probabilities, Bayes factors, and credible intervals. The computational efficiency of this framework stems from exploring segments of the sampling distribution of posterior probabilities for each sample size considered when the conditions for the BvM theorem are satisfied. The numerical studies conducted show that our fast method yields suitable power curve approximation for moderate and large sample sizes. While this method is not appropriate for small sample sizes, it informs practitioners when their required sample sizes are small. More traditional simulation-based design methods can be used in these scenarios since they are often less cumbersome to implement with small sample sizes. 

 Future work could extend this framework to accommodate multiple study objectives. For instance, we may require an $(n, \gamma)$ combination that both satisfies a power criterion and bounds a type I error rate. If the critical value $\gamma$ were chosen algorithmically to bound the type I error rate, we would not be able to use root-finding algorithms as in this paper to select sampling distribution segments. We might also extend this framework for use with more complex models that account for additional covariates in the sample size calculation when comparing two groups. Moreover, the framework presented in the main paper does not support imbalanced two-group sample size determination (i.e., where $n_2 = qn_1$ for some constant $q > 0$). It may be inefficient or impractical to force $q = 1$ when prior information for one group is much more precise, when it is more difficult to sample from one of the groups, or in scenarios where one treatment is much riskier. In Appendix D.2 of the supplement, we extend this framework to settings where practitioners specify this constant $q$.

\section*{Supplementary Material}
These materials include a detailed description of the conditions for Theorem \ref{thm1} and Lemma \ref{lem1} along with their proofs and additional simulation results. The code to implement Algorithm \ref{alg4} and conduct the numerical studies in the main paper is available online: \url{https://github.com/lmhagar/BayesianPower}.

	\section*{Funding Acknowledgement}
	This work was supported by the Natural Sciences and Engineering Research Council of Canada (NSERC) by way of CGS M and PGS D scholarships as well as Grant RGPIN-2019-04212.
	
	
\bibliographystyle{chicago}

\begin{thebibliography}{}

\bibitem[\protect\citeauthoryear{Berry, Carlin, Lee, and Muller}{Berry et~al.}{2011}]{berry2010bayesian}
Berry, S.~M., B.~P. Carlin, J.~J. Lee, and P.~Muller (2011).
\newblock {\em Bayesian Adaptive Methods for Clinical Trials}.
\newblock CRC press.

\bibitem[\protect\citeauthoryear{Brent}{Brent}{1973}]{brent1973algorithm}
Brent, R.~P. (1973).
\newblock An algorithm with guaranteed convergence for finding the minimum of a function of one variable.
\newblock {\em Algorithms for Minimization without Derivatives, Prentice-Hall, Englewood Cliffs, NJ\/}, 61--80.

\bibitem[\protect\citeauthoryear{Brutti, De~Santis, and Gubbiotti}{Brutti et~al.}{2014}]{brutti2014bayesian}
Brutti, P., F.~De~Santis, and S.~Gubbiotti (2014).
\newblock Bayesian-frequentist sample size determination: a game of two priors.
\newblock {\em Metron\/}~{\em 72\/}(2), 133--151.

\bibitem[{Casella et~al.(2004)Casella, Robert, and Wells}]{casella2004generalized}
Casella, G., Robert, C.~P., and Wells, M.~T. (2004).
\newblock Generalized accept-reject sampling schemes.
\newblock {\em Lecture Notes-Monograph Series\/}, 342--347.

\bibitem[\protect\citeauthoryear{Chaloner}{Chaloner}{1996}]{chaloner1996elicitation}
Chaloner, K. (1996).
\newblock Elicitation of prior distributions.
\newblock In {\em Bayesian biostatistics}, pp.\  141--156. Marcel Dekker, New York.

\bibitem[\protect\citeauthoryear{De~Santis}{De~Santis}{2007}]{de2007using}
De~Santis, F. (2007).
\newblock Using historical data for {B}ayesian sample size determination.
\newblock {\em Journal of the Royal Statistical Society: Series A (Statistics in Society)\/}~{\em 170\/}(1), 95--113.

\bibitem[\protect\citeauthoryear{Garthwaite, Kadane, and O'Hagan}{Garthwaite et~al.}{2005}]{garthwaite2005statistical}
Garthwaite, P.~H., J.~B. Kadane, and A.~O'Hagan (2005).
\newblock Statistical methods for eliciting probability distributions.
\newblock {\em Journal of the American Statistical Association\/}~{\em 100\/}(470), 680--701.

\bibitem[\protect\citeauthoryear{Gelman, Carlin, Stern, Dunson, Vehtari, and Rubin}{Gelman et~al.}{2013}]{gelman2013bayesian}
Gelman, A., J.~B. Carlin, H.~S. Stern, D.~B. Dunson, A.~Vehtari, and D.~B. Rubin (2013).
\newblock {\em Bayesian Data Analysis}.
\newblock CRC press.

\bibitem[\protect\citeauthoryear{Gubbiotti and De~Santis}{Gubbiotti and De~Santis}{2011}]{gubbiotti2011bayesian}
Gubbiotti, S. and F.~De~Santis (2011).
\newblock A {B}ayesian method for the choice of the sample size in equivalence trials.
\newblock {\em Australian \& New Zealand Journal of Statistics\/}~{\em 53\/}(4), 443--460.

\bibitem[\protect\citeauthoryear{Hagar and Stevens}{Hagar and Stevens}{2023}]{hagar2023supp}
Hagar, L. and N.~T. Stevens (2024).
\newblock Supplement to ``Fast power curve approximation for posterior analyses".
\newblock \textit{Bayesian Analysis} (submitted).

\bibitem[\protect\citeauthoryear{Hofert and Lemieux}{Hofert and Lemieux}{2020}]{hofert2020package}
Hofert, M. and C.~Lemieux (2020).
\newblock {\em qrng: (Randomized) Quasi-Random Number Generators}.
\newblock R package version 0.0-8.

\bibitem[\protect\citeauthoryear{{\acroauthor{Instituto Nacional de Estadística, Geografía e Informática [National Institute of Statistics, Geography, and Informatics]}{INEGI}}}{{\acroauthor{Instituto Nacional de Estadística, Geografía e Informática [National Institute of Statistics, Geography, and Informatics]}{INEGI}}}{2019}]{enigh2018}
{\acroauthor{Instituto Nacional de Estadística, Geografía e Informática [National Institute of Statistics, Geography, and Informatics]}{INEGI}} (2019).
\newblock Encuesta {N}acional de {I}ngresos y {G}astos de los {H}ogares ({ENIGH}). 2018 {N}ueva serie [{N}ational {S}urvey of {H}ousehold {I}ncome and {E}xpenses. {N}ew edition 2018].
\newblock \url{www.inegi.org.mx/programas/enigh/nc/2018/#Datos_abiertos}.

\bibitem[\protect\citeauthoryear{{\acroauthor{Instituto Nacional de Estadística, Geografía e Informática [National Institute of Statistics, Geography, and Informatics]}{INEGI}}}{{\acroauthor{Instituto Nacional de Estadística, Geografía e Informática [National Institute of Statistics, Geography, and Informatics]}{INEGI}}}{2021}]{enigh2020}
{\acroauthor{Instituto Nacional de Estadística, Geografía e Informática [National Institute of Statistics, Geography, and Informatics]}{INEGI}} (2021).
\newblock Encuesta {N}acional de {I}ngresos y {G}astos de los {H}ogares ({ENIGH}). 2020 {N}ueva serie [{N}ational {S}urvey of {H}ousehold {I}ncome and {E}xpenses. {N}ew edition 2020].
\newblock \url{www.inegi.org.mx/programas/enigh/nc/2020/#Datos_abiertos}.

\bibitem[\protect\citeauthoryear{Jeffreys}{Jeffreys}{1935}]{jeffreys1935some}
Jeffreys, H. (1935).
\newblock Some tests of significance, treated by the theory of probability.
\newblock In {\em Mathematical Proceedings of the Cambridge Philosophical Society}, Volume~31, pp.\  203--222. Cambridge University Press.

\bibitem[\protect\citeauthoryear{Johnson, Tomlinson, Hawker, Granton, and Feldman}{Johnson et~al.}{2010}]{johnson2010methods}
Johnson, S.~R., G.~A. Tomlinson, G.~A. Hawker, J.~T. Granton, and B.~M. Feldman (2010).
\newblock Methods to elicit beliefs for {B}ayesian priors: a systematic review.
\newblock {\em Journal of clinical epidemiology\/}~{\em 63\/}(4), 355--369.

\bibitem[\protect\citeauthoryear{Kass and Raftery}{Kass and Raftery}{1995}]{kass1995bayes}
Kass, R.~E. and A.~E. Raftery (1995).
\newblock Bayes factors.
\newblock {\em Journal of the American Statistical Association\/}~{\em 90\/}(430), 773--795.

\bibitem[\protect\citeauthoryear{Kruschke}{Kruschke}{2018}]{kruschke2018rejecting}
Kruschke, J.~K. (2018).
\newblock Rejecting or accepting parameter values in {B}ayesian estimation.
\newblock {\em Advances in Methods and Practices in Psychological Science\/}~{\em 1\/}(2), 270--280.

\bibitem[\protect\citeauthoryear{Lehmann and Casella}{Lehmann and Casella}{1998}]{lehmann1998theory}
Lehmann, E.~L. and G.~Casella (1998).
\newblock {\em Theory of Point Estimation}.
\newblock Springer Science \& Business Media.

\bibitem[\protect\citeauthoryear{Lemieux}{Lemieux}{2009}]{lemieux2009using}
Lemieux, C. (2009).
\newblock Using quasi--monte carlo in practice.
\newblock In {\em Monte Carlo and Quasi-Monte Carlo Sampling}, pp.\  1--46. Springer.

\bibitem[\protect\citeauthoryear{Morey and Rouder}{Morey and Rouder}{2011}]{morey2011bayes}
Morey, R.~D. and J.~N. Rouder (2011).
\newblock Bayes factor approaches for testing interval null hypotheses.
\newblock {\em Psychological methods\/}~{\em 16\/}(4), 406.

\bibitem[\protect\citeauthoryear{Sobol'}{Sobol'}{1967}]{sobol1967distribution}
Sobol', I.~M. (1967).
\newblock On the distribution of points in a cube and the approximate evaluation of integrals.
\newblock {\em Zhurnal Vychislitel'noi Matematiki i Matematicheskoi Fiziki\/}~{\em 7\/}(4), 784--802.

\bibitem[\protect\citeauthoryear{Spiegelhalter, Abrams, and Myles}{Spiegelhalter et~al.}{2004}]{spiegelhalter2004bayesian}
Spiegelhalter, D.~J., K.~R. Abrams, and J.~P. Myles (2004).
\newblock {\em Bayesian approaches to clinical trials and health-care evaluation}, Volume~13.
\newblock John Wiley \& Sons.

\bibitem[\protect\citeauthoryear{Spiegelhalter, Freedman, and Parmar}{Spiegelhalter et~al.}{1994}]{spiegelhalter1994bayesian}
Spiegelhalter, D.~J., L.~S. Freedman, and M.~K. Parmar (1994).
\newblock Bayesian approaches to randomized trials.
\newblock {\em Journal of the Royal Statistical Society: Series A (Statistics in Society)\/}~{\em 157\/}(3), 357--387.

\bibitem[\protect\citeauthoryear{Stevens and Hagar}{Stevens and Hagar}{2022}]{stevens2022cpm}
Stevens, N.~T. and L.~Hagar (2022).
\newblock Comparative probability metrics: Using posterior probabilities to account for practical equivalence in {A}/{B} tests.
\newblock {\em The American Statistician\/}~{\em 76\/}(3), 224--237.

\bibitem[\protect\citeauthoryear{van~der Vaart}{van~der Vaart}{1998}]{vaart1998bvm}
van~der Vaart, A.~W. (1998).
\newblock {\em Asymptotic Statistics}.
\newblock Cambridge Series in Statistical and Probabilistic Mathematics. Cambridge University Press.

\bibitem[\protect\citeauthoryear{Wang and Gelfand}{Wang and Gelfand}{2002}]{wang2002simulation}
Wang, F. and A.~E. Gelfand (2002).
\newblock A simulation-based approach to bayesian sample size determination for performance under a given model and for separating models.
\newblock {\em Statistical Science\/}~{\em 17\/}(2), 193--208.

\end{thebibliography}

\end{document}


\newcommand{\bb}{\boldsymbol{\beta}}

	\title{Supplementary Material for Fast Power Curve Approximation for Posterior Analyses}

 \author{}


	\date{}

	\maketitle





	\maketitle

	\baselineskip=19.5pt


 	\appendix
\numberwithin{equation}{section}
\renewcommand{\theequation}{\thesection.\arabic{equation}}

\numberwithin{figure}{section}
\renewcommand{\thefigure}{\thesection.\arabic{figure}}

\numberwithin{table}{section}
\renewcommand{\thetable}{\thesection.\arabic{table}}

		\section{Additional Content for Theorem 1}\label{sec:proof}

	\subsection{Conditions for the Bernstein-von Mises Theorem}\label{sec:condBVM}

Theorem 1 from the main text requires that the conditions for the Bernstein-von Mises (BvM) theorem are satisfied for all design distributions that may result from $p_D(\boldsymbol{\eta})$. These conditions are described in more detail in \citet{vaart1998bvm}, starting on page 140. Conditions (B0), (B1), and (B2) concern the likelihood component of the posterior distribution for a parameter $\theta$. (B3) concerns the prior specifications for $\theta$.

\begin{itemize}
    \item [(B0)] The observations are drawn independently and identically from a distribution $P_{\theta_0}$ for some fixed, nonrandom $\theta_0$.
    \vspace*{-2pt}
    \item [(B1)] The parametric statistical model from which the data are generated is differentiable in quadratic mean.  
    \vspace*{-18pt}
    \item [(B2)] There exists a sequence of uniformly consistent tests for testing $H_0: \theta = \theta_0$ against $H_1: \lVert \theta - \theta_0 \rVert \ge \varepsilon$ for every $\varepsilon > 0$.
    \vspace*{-2pt}
    \item [(B3)] Let the prior distribution for $\theta$ be absolutely continuous in a neighbourhood of $\theta_0$ with continuous positive density at $\theta_0$.
\end{itemize}

	\subsection{Conditions for the Asymptotic Normality of the Maximum Likelihood Estimator}\label{sec:MLE}

Theorem 1 from the main text also requires that the design distributions $f(y; \boldsymbol{\eta}_{1,0})$ and $f(y; \boldsymbol{\eta}_{2,0})$ satisfy the regularity conditions for the asymptotic normality of the maximum likelihood estimator. These conditions are detailed in \citet{lehmann1998theory}; they consider a family of probability distributions $\mathcal{P} = \{P_{\theta}: \theta \in \Omega \}$, where $\Omega$ is the parameter space. \citet{lehmann1998theory} use $\theta$ as the unknown parameter with true fixed value $\theta_0$, so we state the conditions using this notation. However, we use $\theta = \theta_1 - \theta_2$ or $\theta = \theta_1/\theta_2$ to compare two characteristics in our framework. For our purposes, the conditions in \citet{lehmann1998theory} must hold for all design distributions (with unknown parameters $\boldsymbol{\eta}_1$ and $\boldsymbol{\eta}_2$ and design values $\boldsymbol{\eta}_{1,0}$ and $\boldsymbol{\eta}_{2,0}$) that could arise from the design prior $p_D(\boldsymbol{\eta})$.  

\citet{lehmann1998theory} detail nine conditions that guarantee the asymptotic normality of the maximum likelihood estimator. We provide the following guidance on where to find more information about these conditions in their text. The first four conditions --  (R0), (R1), (R2), and (R3) -- are described on pages 443 and 444 of their text. (R4) is mentioned as part of Theorem 3.7 on page 447. (R5), (R6), and (R7) are described in Theorem 2.6 on pages 440 and 441. (R8) is mentioned in Theorem 3.10 on page 449.

\begin{itemize}
    \item [(R0)] The distributions $P_{\theta}$ of the observations are distinct.
    \vspace*{-2pt}
    \item [(R1)] The distributions $P_{\theta}$ have common support.
    \vspace*{-2pt}
    \item [(R2)] The observations are $\mathbf{X} = (X_1, ..., X_n)$, where the $X_i$ are identically and independently distributed with probability density function $f(x_i|\theta)$ with respect to a $\sigma$-finite measure $\mu$.
    \vspace*{-2pt}
    \item [(R3)] The parameter space $\Omega$ contains an open set $\omega$ of which the true parameter value $\theta_0$ is an interior point.
    \item [(R4)] For almost all $x$, $f(x|\theta)$ is differentiable with respect to $\theta$ in $\omega$, with derivative $f^{\prime}(x|\theta)$.
    \vspace*{-2pt}
    \item [(R5)] For every $x$ in the set $\{x :  f(x|\theta) > 0 \}$, the density $f(x|\theta)$ is differentiable up to order 3 with respect to $\theta$, and the third derivative is continuous in $\theta$.
    \vspace*{-2pt}
    \item [(R6)] The integral $\int f(x|\theta) d \mu(x)$ can be differentiated three times under the integral sign.
    \vspace*{-2pt}
    \item [(R7)] The Fisher information $\mathcal{I}(\theta)$ satisfies $0 < \mathcal{I}(\theta) < \infty$.
    \vspace*{-2pt}
    \item [(R8)] For any given $\theta_0 \in \Omega$, there exists a positive number $c$ and a function $M(x)$ (both of which may depend on $\theta_0$) such that $\lvert \partial^3 \text{log}f(x|\theta)/\partial \theta^3 \rvert \le M(x)$ for all $\{x :  f(x|\theta) > 0 \}$, $\theta_0 - c < \theta < \theta_0 + c$, and $\mathbb{E}[M(X)] < \infty$.
\end{itemize}

\subsection{Proof of Theorem 1}\label{sec:thm1}

We prove Theorem 1 of the main text in two stages. We begin by proving a simpler version of Theorem 1 where the design prior $p_D(\boldsymbol{\eta})$ is degenerate and assigns a probability of 1 to some $\boldsymbol{\eta}_0$. We first prove part $(a)$ of Theorem 1 under this conditional approach. When data $\boldsymbol{Y}^{_{(n)}}$ are generated, the fraction inside the standard normal CDF of (3.6) in the main text with the posterior approximation method from (3.1) converges to the following normal distribution: 
 \begin{equation}\label{eqn:thm1a}
\sqrt{n}\left(\dfrac{\delta - \theta_0}{\sqrt{\mathcal{I}(\hat\theta_n)^{-1}}} - \dfrac{\hat\theta_n - \theta_0}{\sqrt{\mathcal{I}(\hat\theta_n)^{-1}}}\right) \xrightarrow{d} \mathcal{N}\left(\dfrac{\delta - \theta_0}{\sqrt{\mathcal{I}(\theta_0)^{-1}}}, 1\right).
	\end{equation} 
This result follows by the asymptotic normality of the MLEs $\hat{\boldsymbol{\eta}}_{1,n}$ and $\hat{\boldsymbol{\eta}}_{2,n}$, the continuous mapping theorem because $g(\cdot)$ and $h(\cdot)$ are differentiable at the design values, and Slutsky's theorem since $\mathcal{I}(\hat\theta_n)^{-1} \xrightarrow{P} \mathcal{I}(\theta_0)^{-1}$. When psuedorandom sequences $\boldsymbol{U} \overset{\text{i.i.d.}}{\sim} \mathcal{U}([0,1]^{2d})$ are input into Algorithm 1, the left side of (\ref{eqn:thm1a}) follows the normal distribution on the right side exactly. We obtain the result in part $(a)$ under the conditional approach by a second application of the continuous mapping theorem with the function $\Phi(\cdot)$. 

To prove part $(b)$ under the conditional approach, we note that the approximations in (3.1) and (3.2) of the main text are virtually the same as $n \rightarrow \infty$. Under the conditions for Theorem 1, the posterior mode $\tilde{ \boldsymbol{\eta}}_{j,n}$ converges in probability to $\boldsymbol{\eta}_{j,0}$ for $j = 1,2$. The following result also holds for $\mathcal{J}_j(\tilde{\boldsymbol{\eta}}_{j,n})/n$ in (3.2) of the main text: 
\begin{equation}\label{eqn:fish_limit}
    \dfrac{1}{n}\mathcal{J}_j(\tilde{\boldsymbol{\eta}}_{j,n}) = \left[-\dfrac{1}{n}\sum_{i = 1}^n\dfrac{\partial^2}{\partial\boldsymbol{\eta}_j^2}\text{log}(f(y_{ij}; \boldsymbol{\eta}_j)) - \dfrac{1}{n}\dfrac{\partial^2}{\partial\boldsymbol{\eta}_j^2}\text{log}(p_j(\boldsymbol{\eta}_j))\right]_{\boldsymbol{\eta}_j = \tilde{ \boldsymbol{\eta}}_{j,n}} \xrightarrow{P} \mathcal{I}(\boldsymbol{\eta}_{j,0}).
\end{equation}
Because $\tilde{\boldsymbol{\eta}}_{j,n} - \hat{\boldsymbol{\eta}}_{j,n} \xrightarrow{P} 0$, the mean and variance of the normal distribution in (3.2) of the main text respectively approximate $\hat\theta_n$ and $\mathcal{I}(\hat{{\theta}}_n)^{-1}/n$ in (3.1) for large sample sizes $n$ by the continuous mapping theorem. Under the conditional approach, the result in part $(b)$ then follows from part $(a)$. 

We use this result for the conditional approach to prove Theorem 1 for the predictive approach with nondegenerate design priors $p_D(\boldsymbol{\eta})$. The conditions for Theorem 1 ensure that the previous result holds for all $\boldsymbol{\eta}_0$ such that $p_D(\boldsymbol{\eta}_0) > 0$. Theorem 1 assumes that $\boldsymbol{\eta}_0 \sim p_D(\boldsymbol{\eta})$. For nondegenerate $p_D(\boldsymbol{\eta})$, we must integrate with respect to $\boldsymbol{\eta}$: Theorem 1 also holds true for the predictive approach by yet another application of the continuous mapping theorem. $\qed$ 



\section{Proof of Lemma 1}

We prove part $(a)$ of Lemma 1 from the main text in three steps. In the first step, we prove that Algorithm 1 prompts 
\begin{equation}\label{eqn:lem1a1}
\hat{\boldsymbol{\eta}}_{1,n}(\boldsymbol{u}_r)_{[k]} = \boldsymbol{\eta}_{1,0 [k]} +\dfrac{\omega_k(u_1, ..., u_k)}{\sqrt{n}} ~~~\text{and}~~~ \hat{\boldsymbol{\eta}}_{2,n}(\boldsymbol{u}_r)_{[k]} = \boldsymbol{\eta}_{2,0 [k]} +\dfrac{\omega_{d+k}(u_{d+1}, ..., u_{d+k})}{\sqrt{n}},
  \end{equation} 
  for $k = 1, ..., d$ such that $\omega_k(\cdot)$ and $\omega_{d + k}(\cdot)$ are functions that do not depend on the sample size. In (\ref{eqn:lem1a1}), we first consider $\boldsymbol{u} = (u_1, u_2, ..., u_{2d}) \in [0,1]^{2d}$ because points of this dimension are input into Algorithms 1, 2, and 3.

We only present the proof of (\ref{eqn:lem1a1}) for group 1 since the proof for group 2 follows the same process. We use induction on the dimension $d$ of $\boldsymbol{\eta}_{1}$ for this proof. We show the base case corresponding to a model with $d = 2$. To simplify notation, we let
 \begin{equation*}\label{eqn:fish2}
\mathcal{I}(\boldsymbol{\eta}_{1,0})^{-1} = \begin{bmatrix}
    \sigma^2_{11}       & \rho_{12}\sigma_{11}\sigma_{22} \\
    \rho_{12}\sigma_{11}\sigma_{22}       & \sigma^2_{22} 
\end{bmatrix}.
	\end{equation*} 
By properties of the bivariate conditional normal distribution, it follows that
 \begin{equation}\label{eqn:dim1}
\hat{\boldsymbol{\eta}}_{1,n}(\boldsymbol{u}_r)_{[1]} = \boldsymbol{\eta}_{1,0 [1]} + \dfrac{1}{\sqrt{n}}\Phi^{-1}(u_1)\sigma_{11} ~~~\text{and}
	\end{equation} 
  \begin{equation}\label{eqn:dim2}
\hat{\boldsymbol{\eta}}_{1,n}(\boldsymbol{u}_r)_{[2]} = \boldsymbol{\eta}_{1,0 [2]} + \dfrac{1}{\sqrt{n}}\sigma_{22}\left[\Phi^{-1}(u_1)\rho_{12} + \Phi^{-1}(u_2)\sqrt{1 - \rho^2_{12}}\right].
	\end{equation}
 As in the main text, the square brackets in the subscripts denote the indices of the relevant subvectors. The result in (\ref{eqn:lem1a1}) therefore holds true when $d = 2$, where $\omega_1(u_1)$ and $\omega_2(u_1, u_2)$ are given by the expressions to the right of the $1/\sqrt{n}$ terms in (\ref{eqn:dim1}) and (\ref{eqn:dim2}), respectively.

 For the inductive hypothesis, we assume that the result in (\ref{eqn:lem1a1}) holds true for a model with $d = l$ parameters. For the inductive conclusion, we show that this implies the result also holds for a model with $d = l + 1$ parameters. Because $\hat{\boldsymbol{\eta}}_{1,n}(\boldsymbol{u}_r)_{[1]}, \dots, \hat{\boldsymbol{\eta}}_{1,n}(\boldsymbol{u}_r)_{[l]}$ only depend on the components with smaller indices, we just need to prove that the result in (\ref{eqn:lem1a1}) holds for $\hat{\boldsymbol{\eta}}_{1,n}(\boldsymbol{u}_r)_{[l+1]}$. That result in conjunction with the inductive hypothesis proves the inductive conclusion. To prove the inductive conclusion, we introduce the following block matrix notation:
  \begin{equation*}\label{eqn:fishlp1}
\mathcal{I}(\boldsymbol{\eta}_{1,0})^{-1} = \begin{bmatrix}
    \boldsymbol{\Sigma}_{l, l}       & \boldsymbol{\Sigma}_{l, 1} \\
    \boldsymbol{\Sigma}_{1, l}       & \Sigma_{l+1, l+1} 
\end{bmatrix},
	\end{equation*} 
 where $\boldsymbol{\Sigma}_{l,l}$ is a $l \times l$ matrix, $\boldsymbol{\Sigma}_{l, 1}$ is a $l \times 1$ matrix, $\boldsymbol{\Sigma}_{1, l} = \boldsymbol{\Sigma}_{l, 1}^T$, and $\Sigma_{l + 1, l + 1}$ is scalar.

 The marginal distribution of $\hat{\boldsymbol{\eta}}_{1,n [l+1]}$ conditional on the previously generated \linebreak $\hat{\boldsymbol{\eta}}_{1,n}(\boldsymbol{u}_r)_{[k]}$ for $k = 1, ..., l$ is
 \begin{equation*}\label{eqn:dimlpl1}
 \mathcal{N}\left(\boldsymbol{\eta}_{1,0 [l+1]} + \dfrac{1}{\sqrt{n}}\boldsymbol{\Sigma}_{1, l}\boldsymbol{\Sigma}_{l, l}^{-1}\begin{pmatrix}
    \omega_1(u_1)  \\ \vdots \\
    \omega_l(u_1, ..., u_l) 
\end{pmatrix}, \dfrac{1}{n}\left[\Sigma_{l + 1, l + 1} - \boldsymbol{\Sigma}_{1, l}\boldsymbol{\Sigma}_{l, l}^{-1}\boldsymbol{\Sigma}_{l, 1}\right] \right).
\end{equation*}
Therefore, we have that 
  \begin{equation}\label{eqn:dimlp1}
  \begin{split}
\hat{\boldsymbol{\eta}}_{1,n}(\boldsymbol{u}_r)_{[l+1]} = \boldsymbol{\eta}_{1,0[l+1]} &+ \dfrac{1}{\sqrt{n}}\boldsymbol{\Sigma}_{1, l}\boldsymbol{\Sigma}_{l, l}^{-1}\begin{pmatrix}
    \omega_1(u_1)  \\ \vdots \\
    \omega_l(u_1, ..., u_l) 
\end{pmatrix} \\ &+ \dfrac{1}{\sqrt{n}}\Phi^{-1}(u_{l+1})\left[\Sigma_{l + 1, l + 1} - \boldsymbol{\Sigma}_{1, l}\boldsymbol{\Sigma}_{l, l}^{-1}\boldsymbol{\Sigma}_{l, 1}\right].
\end{split}
	\end{equation}
 The result from (\ref{eqn:lem1a1}) holds for $\hat{\boldsymbol{\eta}}_{1,n}(\boldsymbol{u}_r)_{[l+1]}$ if we take $\omega_{l + 1}(u_1, ..., u_{l + 1})$ as the sum of the two components to the right of the $1/\sqrt{n}$ terms in (\ref{eqn:dimlp1}). By mathematical induction, (\ref{eqn:lem1a1}) is true for an arbitrary model with $d$ parameters.

 The second step to prove part $(a)$ of Lemma 1 involves showing that 
  \begin{equation}\label{eqn:lem1a2}
h(g(\hat{\boldsymbol{\eta}}_{1,n}(\boldsymbol{u}_r)), g(\hat{\boldsymbol{\eta}}_{2,n}(\boldsymbol{u}_r))) \approx h(g(\boldsymbol{\eta}_{1,0}), g(\boldsymbol{\eta}_{2,0})) +\dfrac{\omega_{\dagger}(u_{1}, ..., u_{2d})}{\sqrt{n}},
  \end{equation}
for sufficiently large $n$, where $\omega_{\dagger}(\cdot)$ is a function that does not depend on $n$. The result in (\ref{eqn:lem1a2}) follows from the first-order Taylor expansion of $h(g(\hat{\boldsymbol{\eta}}_{1,n}(\boldsymbol{u}_r)),$ $ g(\hat{\boldsymbol{\eta}}_{2,n}(\boldsymbol{u}_r)))$ around $(\boldsymbol{\eta}_{1,0}, \boldsymbol{\eta}_{2,0})$. We have that 
 \begin{equation}\label{eqn:taylorg}
   \begin{split}
& h(g(\hat{\boldsymbol{\eta}}_{1,n}(\boldsymbol{u}_r)), g(\hat{\boldsymbol{\eta}}_{2,n}(\boldsymbol{u}_r))) - h(g(\boldsymbol{\eta}_{1,0}), g(\boldsymbol{\eta}_{2,0})) \\ &\approx \sum_{j = 1}^2\sum_{k = 1}^d\dfrac{\partial h}{\partial g_j}\dfrac{\partial g_j}{\partial \boldsymbol{\eta}_{j [k]}}\bigg|_{(\boldsymbol{\eta}_1, \boldsymbol{\eta}_2)  = (\boldsymbol{\eta}_{1,0}, \boldsymbol{\eta}_{2,0})}\left[\hat{\boldsymbol{\eta}}_{j,n}(\boldsymbol{u}_r)_{[k]} -  \boldsymbol{\eta}_{j,0[k]}\right] \\
&\approx \dfrac{1}{\sqrt{n}}\left[\sum_{j = 1}^2 \sum_{k = 1}^d\dfrac{\partial h}{\partial g_j}\dfrac{\partial g_j}{\partial \boldsymbol{\eta}_{j [k]}}\bigg|_{(\boldsymbol{\eta}_1, \boldsymbol{\eta}_2)  = (\boldsymbol{\eta}_{1,0}, \boldsymbol{\eta}_{2,0})} \omega_{(j-1)d + k}(u_{(j-1)d + 1}, ..., u_{(j-1)d + k}) \right].
\end{split}
\end{equation}
The result from (\ref{eqn:lem1a2}) follows if we let $\omega_{\dagger}(\cdot)$ be the sum on the right side of the $1/\sqrt{n}$ term in (\ref{eqn:taylorg}). However, this Taylor series expansion may only be suitable for large sample sizes $n$ -- when $\hat{\boldsymbol{\eta}}_{j,n}(\boldsymbol{u}_r)$ is sufficiently near ${\boldsymbol{\eta}}_{j,0}$ for $j = 1, 2$. 


To prove part $(a)$ of Lemma 1, the final step of the proof involves showing that the following result holds for sufficiently large $n$:
 \begin{equation}\label{eqn:Z}
   \begin{split}
 \dfrac{\delta - \underline\theta_{r}^{_{(n)}}}{\sqrt{\underline{\tau}_{r}^{_{(n)}}}} & \approx \dfrac{\delta - (h(g(\boldsymbol{\eta}_{1,0}), g(\boldsymbol{\eta}_{2,0})) + \omega_{\dagger}(u_1, ..., u_{2d})/\sqrt{n})}{\sqrt{\mathcal{I}(\theta_0)^{-1}/n}} \\
& = \dfrac{\delta - \theta_0}{\sqrt{\mathcal{I}(\theta_0)^{-1}}}\sqrt{n} - \dfrac{\omega_{\dagger}(u_1, ..., u_{2d})}{\sqrt{\mathcal{I}(\theta_0)^{-1}}}.
\end{split}
\end{equation}
 The approximate equivalence of the numerators in the first line of (\ref{eqn:Z}) follows from (\ref{eqn:lem1a2}) and the fact that $\tilde{\boldsymbol{\eta}}_{j,n} - \hat{\boldsymbol{\eta}}_{j,n}$ and ${\boldsymbol{\eta}}_{j,n}^* - \hat{\boldsymbol{\eta}}_{j,n}$ converge in probability to 0 as mentioned in the main text. Moreover, $\underline{\tau}_{r}^{_{(n)}} \approx \mathcal{I}(\theta_0)^{-1}/n$ for sufficiently large $n$ by the continuous mapping theorem for Algorithm 1, (\ref{eqn:fish_limit}) for Algorithm 2, and similar logic to (\ref{eqn:fish_limit}) for Algorithm 3. The second line of (\ref{eqn:Z}) holds because $\theta_0 = h(g(\boldsymbol{\eta}_{1,0}), g(\boldsymbol{\eta}_{2,0}))$. The expression in (\ref{eqn:Z}) takes the form $a(\delta, \theta_0)\sqrt{n} + b(\boldsymbol{u}_r)$ since neither fraction in the second line depends on $n$. 
 
 Unlike (\ref{eqn:lem1a1}), (\ref{eqn:lem1a2}), and (\ref{eqn:Z}), part $(a)$ of Lemma 1 considers points $\boldsymbol{u_r} \in [0,1]^{2d + 1}$. For a given point $\boldsymbol{u}_r$ input into Algorithm 1, 2, or 3, $\boldsymbol{\eta}_0$ is indexed by the final coordinate of $\boldsymbol{u}_r$. Thus, $\theta_0$ is a function of $u_{2d+1}$ and $b(\cdot)$ is technically only a function of $u_1, ..., u_{2d}$. Part $(a)$ of Lemma 1 follows by using the normal CDF as in (3.6) of the main text. We note that the function $a(\delta, \theta_0)$, which is the fraction to the left of the $\sqrt{n}$ term in the second line of (\ref{eqn:Z}), must incorporate monotonic transformations applied to the posterior of $\theta$ to improve the suitability of its normal approximation.


 Part $(b)$ of Lemma 1 follows from taking the derivative of the approximation to $p^{\delta_U}_{n, \boldsymbol{u}_r, \zeta} - p^{\delta_L}_{n, \boldsymbol{u}_r, \zeta}$ prompted by part $(a)$ with respect to the sample size $n$:
  \begin{equation}\label{eqn:deriv_phi}
   \begin{split}
 & \dfrac{d}{dn}\left[\ p^{\delta_U}_{n, \boldsymbol{u}_r, \zeta} - p^{\delta_L}_{n, \boldsymbol{u}_r, \zeta}\right] \\ & \approx \dfrac{d}{dn}\left[\ \Phi\left(a(\delta_U, \theta_0)\sqrt{n} + b(\boldsymbol{u}_r)\right) - \Phi\left(a(\delta_L, \theta_0)\sqrt{n} + b(\boldsymbol{u}_r)\right)\right] \\
& = \dfrac{a(\delta_U, \theta_0)\phi\left(a(\delta_U, \theta_0)\sqrt{n} + b(\boldsymbol{u}_r)\right) - a(\delta_L, \theta_0)\phi\left(a(\delta_L, \theta_0)\sqrt{n} + b(\boldsymbol{u}_r)\right)}{2\sqrt{n}},
\end{split}
\end{equation}
where $\phi(\cdot)$ is the probability density function of the standard normal distribution. This derivative must be positive for sufficiently large $n$ where the approximation from part $(a)$ of Lemma 1 holds. When $\theta_0 \in (\delta_L, \delta_U)$, $a(\delta_U, \theta_0)$ is positive and $a(\delta_L, \theta_0)$ is negative. Because the normal distribution takes support over the entire real line, $\phi(\cdot)$ returns a positive value for any real input. If $\delta_L$ or $\delta_U$ is not finite, then its component of the difference in the numerator of (\ref{eqn:deriv_phi}) is zero. The remaining component of the numerator is still positive and so is the derivative in (\ref{eqn:deriv_phi}). Hence, part $(b)$ of Lemma 1 is true, and $p^{\delta_U}_{n, \boldsymbol{u}_r, \zeta} - p^{\delta_L}_{n, \boldsymbol{u}_r, \zeta}$ is an increasing function for sufficiently large $n$. $\qed$




\section{Visualization of Computational Efficiency}


The motivating example with gamma tail probabilities from Section 2 of the main text considers the posteriors of $\boldsymbol{\eta}_j = (\alpha_j, \lambda_j)$ for $j = 1, 2$. That example therefore has a dimension of $2d = 4$ or $2d + 1 = 5$ depending on whether we consider the conditional or predictive approach. Here, we consider a two-dimensional example involving the comparison of Bernoulli proportions that is easier to visualize. Even though this model has a conjugate prior, it is considered for illustrative purposes. We let $\theta_j$ be the Bernoulli success probability for model $j = 1, 2$. For this example, we let $\eta_j = \text{log}(\theta_j) - \text{log}(1 - \theta_j)$ to improve the quality of the normal approximations to the posterior for $\eta_j$ and the sampling distribution of its MLE. This simple transformation is based on the canonical form of the standard Bernoulli model \citep{lehmann1998theory}.

We compare the Bernoulli probabilities via their difference: $\theta_1 - \theta_2$. Because $\theta_1 - \theta_2 \in (-1, 1)$, we similarly consider the posterior distribution of $\text{log}(\theta_1 - \theta_2 + 1) - \text{log}(1 - \theta_1 + \theta_2)$ to improve the quality of the normal approximations for moderate $n$. The selected transformation for $\theta$ is a straightforward generalization of the transformation applied to each $\eta_j$ parameter. An alternative monotonic transformation could have been chosen to minimize the Kullback-Leibler divergence \citep{gelman2013bayesian} for the normal approximation to the posterior. However, the optimal transformation likely depends on the sample size and the data generated from the prior predictive distributions. Any transformations such that $\boldsymbol{\eta}_1$ and $\boldsymbol{\eta}_2$ have support over $\mathbb{R}^{d}$ and $\theta$ has support over the entire real line are suitable in the limiting case. To streamline our methods, we suggest using visualization techniques to compare the suitability of several candidate monotonic transformations if necessary.

We consider the conditional approach to specify $p_D(\boldsymbol{\eta})$ to reduce the dimension of the simulation for visualization purposes. With this approach, we choose design values of $\theta_{1,0} = 0.15$ and $\theta_{2,0} = 0.14$, which gives rise to design values $\eta_{1, 0}$ and $\eta_{2,0}$ on the logit scale. We specify informative analysis priors for the Bernoulli parameters: $\text{BETA}(3.75, 21.25)$ for $\theta_1$ and $\text{BETA}(3.50, 21.50)$ for $\theta_2$. These beta distributions have modes that roughly align with the design values $\theta_{1, 0}$ and $\theta_{2, 0}$, and these beta priors induce priors on the variables $\eta_1$ and $\eta_2$. We consider power curve approximation for analyses with posterior probabilities, where $(\delta_L, \delta_U) = (-0.05, 0.05)$ on the probability scale, $\gamma = 0.8$, and $\Gamma = 0.6$. For concision, we only consider the method to map posteriors to $[0,1]^{2d}$ from Algorithm 2 of the main text with $m = 1024$. 

\begin{figure}[!tb] \centering 
		\includegraphics[width = 0.95\textwidth]{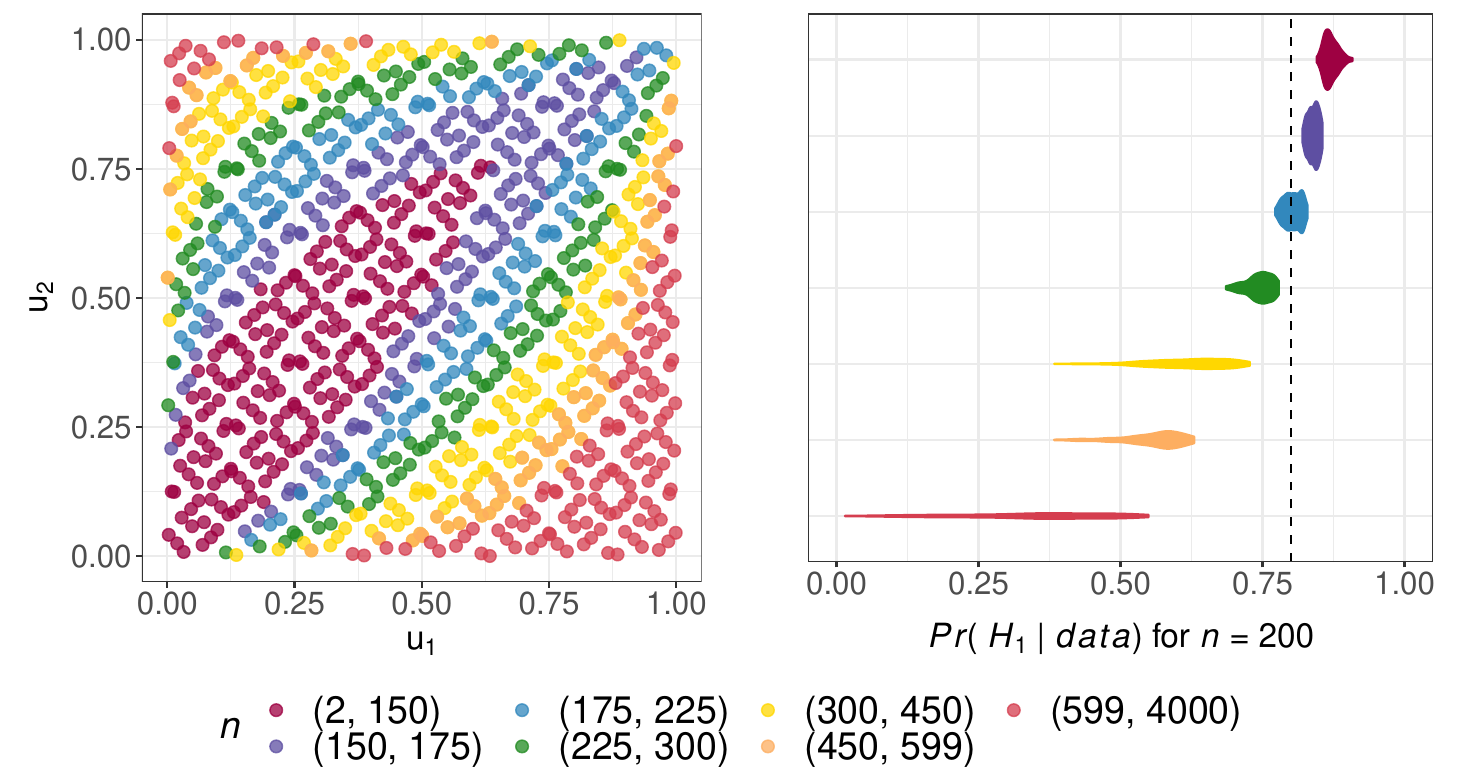} 
		\caption{\label{fig:points2} Left: Visualization of which points in $[0,1]^2$ are used to explore at least one $n$ value in the various sample size ranges via the root-finding algorithm. Right: Violin plots for segments of the sampling distribution of posterior probabilities when $n = 200$; the vertical line is at $\gamma = 0.8$.} 
	\end{figure}

The left plot of Figure \ref{fig:points2} decomposes the results of the root-finding algorithm for one approximated power curve for this Bernoulli example. For instance, only the pink points in the lower left corner assessed power for at least one sample size $n \in (2, 150)$ when input into the root-finding procedure. We note that in most iterations of the root-finding algorithm, the sample size $n$ is noninteger, which does not present issues for our method. This targeted exploration approach allows us to prioritize segments of the sampling distribution of posterior probabilities based on the sample size $n$. Only the blue points from $[0, 1]^2$ considered at least one sample size $n \in (175, 225)$. These are the points for which the posterior probability in (1.3) of the main text is close to the critical value $\gamma = 0.8$. For reference, the final sample size recommendation for this example was $\lceil n^* \rceil = 269$.

The right plot of Figure \ref{fig:points2} visualizes segments from the sampling distribution of posterior probabilities for $n = 200$ conditional on the categorizations from the left plot. It is wasteful to use the pink points to consider $n \in (175, 225)$ because those points satisfy $p^{\delta_U}_{n, \boldsymbol{u}_r, \zeta} - p^{\delta_L}_{n, \boldsymbol{u}_r, \zeta} = \gamma$ for some sample size $n \in (2, 150)$. Based on Lemma 1 from the main text, the posterior probabilities for the pink points with $n \in (175, 225)$ should therefore be much larger than $\gamma$. This result is confirmed by comparing the pink density to the dotted vertical line at $\gamma = 0.8$. By similar logic, it is wasteful to consider the red points for $n \in (175, 225)$ since the posterior probabilities for those points will be much smaller than $\gamma$.


 In Figure \ref{fig:points2}, the sample size categories were created to exclude most sample sizes from the first few iterations of the root-finding algorithm. This categorization limits the number of points in $[0,1]^2$ that belong to more than one of the seven categories for clearer visualization. We emphasize that these categories are not used in our power curve approximation method. Instead, they illustrate the targeted nature of how we explore the approximate sampling distribution of posterior probabilities for each sample size $n$ considered.

\begin{figure}[!t] \centering 
		\includegraphics[width = 0.7\textwidth]{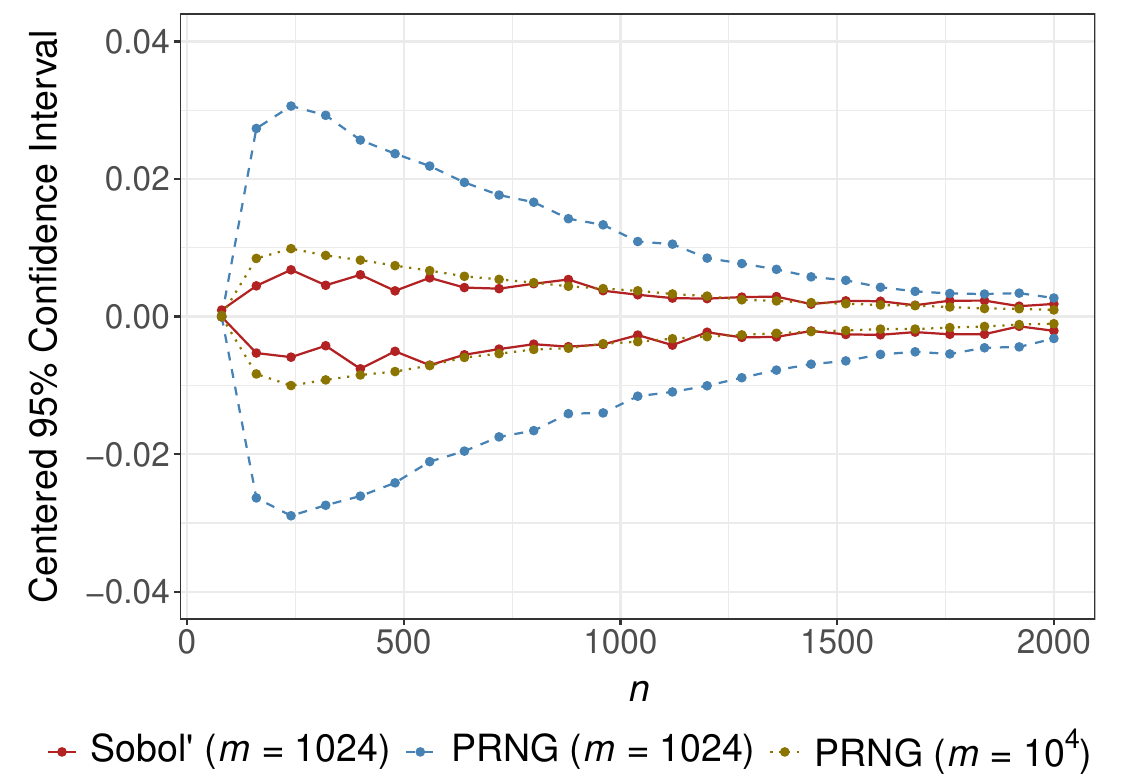} 
		\caption{\label{fig:power.q} Endpoints of the centered 95\% confidence intervals for power obtained with Sobol' and pseudorandom (PRNG) sequences of various lengths.} 
	\end{figure}

We now assess the impact of using Sobol' sequences with our power curve approximation method. To do so, we approximated 1000 power curves for this Bernoulli example using Algorithm 4 from the main text with sequences from a pseudorandom number generator of length $m = 1024$. As before, only the method for posterior mapping in Algorithm 2 was considered. We then repeated this process using Algorithm 4 with Sobol' sequences of length $m = 1024$.  We used the 1000 power curves corresponding to each sequence type (Sobol' and pseudorandom) to estimate power for the following sample sizes: $n =\{80, 160, ..., 2000\}$. For each sample size and sequence type, we obtained a 95\% confidence interval for power using the percentile bootstrap method \citep{efron1982jackknife}. We then created centered confidence intervals by subtracting the mean of the 1000 power estimates from each confidence interval endpoint. Figure \ref{fig:power.q} depicts these results for the sample sizes $n$ and two sequence types considered. 

 Figure \ref{fig:power.q} illustrates that the Sobol' sequence gives rise to much more precise power estimates than pseudorandom sequences -- particularly for sample sizes where power is not near 0 or 1. We repeated the process detailed in the previous paragraph to generate 1000 power curves via Algorithm 4 with pseudorandom sequences of length $m = 10^4$. The power estimates obtained using Sobol' sequences with length $m = 1024$ are roughly as precise as those obtained with pseudorandom sequences of length $m = 10^4$. Similar results were observed for more extensive numerical studies. Using Sobol' sequences therefore allows us to estimate power with the same precision using approximately an order of magnitude fewer points. 
 

 To conclude this section, we concretely detail the magnitude of the gains in computational efficiency that are attributable to the use of (i) sampling distribution segments, (ii) Sobol' sequences, and (iii) both. The following runtimes for this Bernoulli example were measured on a standard laptop without parallelization. When using Algorithm 4 from the main text with Sobol' sequences of length $m = 1024$, it took just under 0.3 seconds to approximate the power curve. The 0.99-quantile of this power curve is roughly $n = 1620$. It took about 25 seconds to construct a power curve by obtaining power estimates with a single Sobol' sequence ($m = 1024$) at $n = \{2, 3, \dots, 1620 \}$. For this example, using sampling distribution segments to approximate the power curve is roughly 83 times more computationally efficient than considering entire sampling distributions. That is, we approximate the power curve $(83-1)\times100 = 8200$\% faster.
 
 When using Algorithm 4 with pseudorandom sequences of length $m = 10^4$ as informed by Figure \ref{fig:power.q}, it took roughly 2.6 seconds to approximate the power curve. These pseudorandom sequences are 9.77 times longer than the Sobol' sequences considered in the previous paragraph. Yet, the use of Sobol' sequences with this example only reduces the runtime by a factor of roughly 9 (or by roughly 800\%) since there is computational overhead associated with choosing the initial sample size $n_0$ in Line 5 of Algorithm 4. Moreover, it took just over 4 minutes to construct a power curve by obtaining power estimates with a single pseudorandom sequence ($m = 10^4$) at $n = \{2, 3, \dots, 1620 \}$. We therefore reduced the runtime by a factor of 800 when combining the use of sampling distribution segments and Sobol' sequences. 
 

 We emphasize that the gains in computational efficiency detailed above are specific to the Bernoulli example with the conditional approach considered in this subsection. In general, the extent of the computational savings depends on the statistical models, design inputs, and the magnitude of the sample sizes that correspond to high study power. Furthermore, this discussion surrounding efficiency gains did not account for the computational savings that arise from using analytical posterior approximation instead of computational approximation methods. 


\section{Additional Numerical Studies}

		\subsection{Numerical Studies with Bayes Factors}\label{sec:poni}

 		\begin{figure}[!b] \centering 
		\includegraphics[width = 0.95\textwidth]{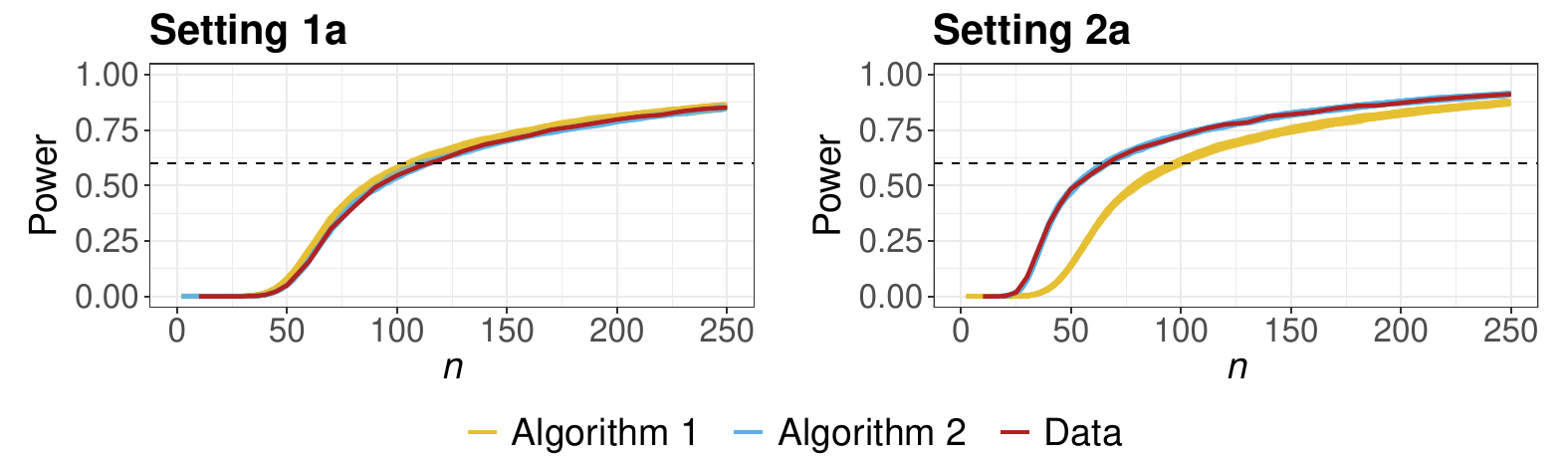} 
		\caption{\label{fig:power1} 100 power curves obtained via Algorithms 1 (yellow) and 2 (blue), power curve estimated via simulated data (red), and target power $\Gamma$ (dotted line) for Settings 1a and 2a with hypothesis tests facilitated via NOH Bayes factors.} 
	\end{figure}

      We now compare the performance of our power curve approximation procedure for several posterior analyses with Bayes factors. We only consider the conditional approach here for concision. To do so, we modify Settings 1a and 2a from Section 5.1 of the main text. We do not change the target power, interval $(\delta_L, \delta_U)$, design values, or analysis priors. For Setting 1a with the uninformative analysis priors, $Pr(\delta_L < \theta_1/\theta_2 < \delta_U) = 0.0128$. We consider a threshold for the nonoverlapping-hypotheses (NOH) Bayes factor of $K = 100$ for illustration. By (1.4) of the main text, this corresponds to a critical value of $\gamma = 0.5652$. For Setting 1b with the informative analysis priors, $Pr(\delta_L < \theta_1/\theta_2 < \delta_U) = 0.2835$. We consider a threshold for the NOH Bayes factor of $K = 3$ for illustration, which corresponds to a critical value of $\gamma = 0.5428$. This example illustrates the importance of considering the impact of the analysis prior for $\theta$ when choosing a threshold $K$. 
      
      The numerical study in this subsection was otherwise carried out using the same process as described in Section 5.1 of the main text. For each setting, we obtained 100 approximations to the power curve using Algorithm 4 with $\zeta = \{\text{Alg. 1}, \text{Alg. 2}\}$. The results for Settings 1a (left) and 2a (right) are depicted in Figure \ref{fig:power1}. This numerical study supports similar conclusions as those drawn regarding Settings 1a and 2a for hypothesis tests with posterior probabilities in Figure 2 of the main text.

\subsection{Numerical Studies with Imbalanced Sample Sizes}\label{sec:num.q}

In Section 6 of the main text, we acknowledge that our framework as presented in the main paper does not support imbalanced sample size determination (i.e., where $n_2 = qn_1$ for some constant $q > 0$). In this subsection, we describe how to extend our methods to allow for imbalanced sample size determination. This procedure requires practitioners to choose the constant $q$ a priori. When $n_1 \ne n_2$, we use the following limiting posteriors for each group: $\mathcal{N}(\boldsymbol{\eta}_{1,0}, \mathcal{I}(\boldsymbol{\eta}_{1,0})^{-1}/n)$ and $\mathcal{N}(\boldsymbol{\eta}_{2,0}, \mathcal{I}(\boldsymbol{\eta}_{2,0})^{-1}/(qn))$. To apply the multivariate delta method to obtain the limiting posterior of $\theta = h(g(\boldsymbol{\eta}_{1}), g(\boldsymbol{\eta}_{2}))$, both the limiting variances of $\boldsymbol{\eta}_{1}$ and $\boldsymbol{\eta}_2$ must be functions of $n$. We therefore treat $\mathcal{I}(\boldsymbol{\eta}_{2,0})^{-1}/q$ as the inverse Fisher information for $\boldsymbol{\eta}_2$ evaluated at the design value $\boldsymbol{\eta}_{2,0}$. This modification is also incorporated into the process to simulate maximum likelihood estimates for $\boldsymbol{\eta}_2$ in Algorithms 1, 2, and 3 of the main text. That is, the variability in the marginal limiting distribution of $\hat{\boldsymbol{\eta}}_{2,qn}$ is scaled to reflect the larger ($q > 1$) or smaller ($0 < q < 1$) sample size in the second group. Similar modifications are also made when taking the normal approximation to the posterior of $\theta$ in (3.1), (3.2), and (3.5) of the main text. No other modifications are required to account for imbalanced sample sizes.

  		\begin{figure}[!b] \centering 
		\includegraphics[width = 0.95\textwidth]{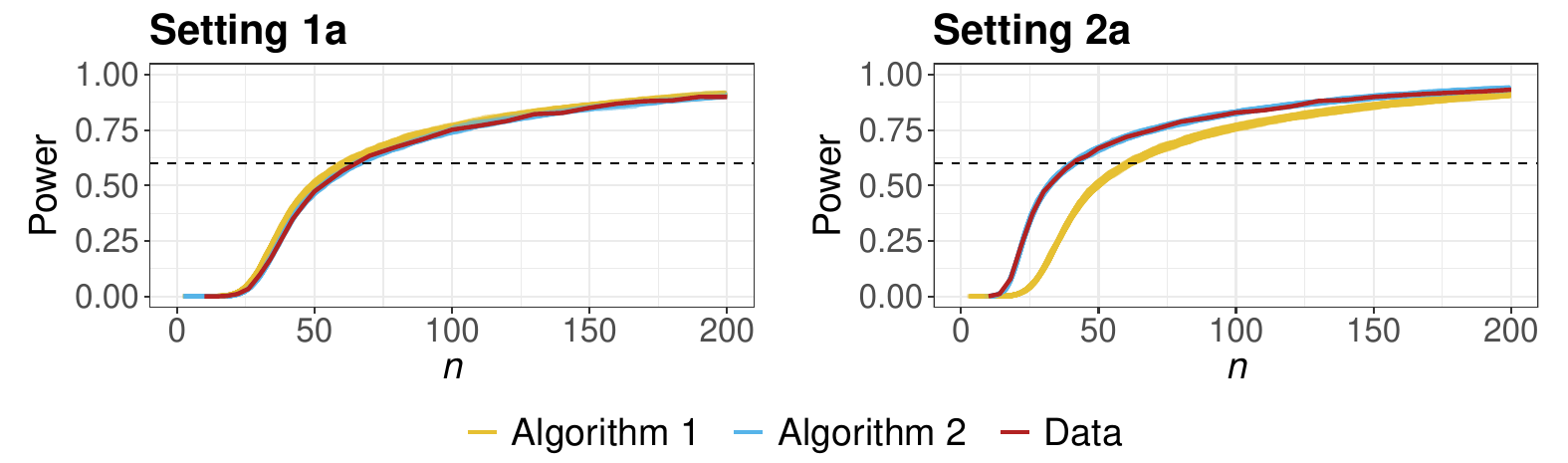} 
		\caption{\label{fig:power3} 100 power curves obtained via Algorithms 1 (yellow) and 2 (blue), power curve estimated via simulated data (red), and target power $\Gamma$ (dotted line) for Settings 1a and 2a with hypothesis tests with imbalanced sample sizes.} 
	\end{figure}

   Lastly, we evaluate the performance of our power curve approximation procedure with several scenarios that have imbalanced sample sizes. We reuse Settings 1a and 2a with the conditional approach from Section 5.1 of the main text. The only differences between this numerical study and the one conducted in Section 5.1 are those described in the previous paragraphs. We choose $q = 2$ (i.e., $n_2 =2n_1)$ for this numerical study. This reflects the male ($j = 2$) provider group having more observations than the female ($j = 1$) one in the motivating example from Section 2 of the main text. 	For each setting, we obtained 100 approximations to the power curve using Algorithm 4 with $\zeta = \{\text{Alg. 1}, \text{Alg. 2}\}$. These results are depicted in Figure \ref{fig:power3}. We again reach similar conclusions as those drawn for Settings 1a and 2a using Figure 2 of the main text.
	
	
\bibliographystyle{chicago}